\documentclass[%
 reprint,
 amsmath,amssymb,
 aps,
 pra,
floatfix,
superscriptaddress
]{revtex4-2}

\usepackage{comment}
\usepackage{graphicx}
\usepackage{dcolumn}
\usepackage{bm}
\usepackage[normalem]{ulem}
\usepackage[
range-units = single,
]{siunitx}
\usepackage{xspace}
\usepackage{xcolor}
\usepackage[colorlinks=true, allcolors=blue]{hyperref}

\usepackage{xstring}

\makeatletter
\newcommand*{\rom}[1]{\expandafter\@slowromancap\romannumeral #1@}
\makeatother

\usepackage{xcolor}
\definecolor{lightblue}{rgb}{0., .7, 0.}

\newcommand*{\aref}[1]{%
	\IfBeginWith{#1}{eq:}{Eq.~\eqref{#1}}{}%
	\IfBeginWith{#1}{fig:}{Fig.~\ref{#1}}{}%
	\IfBeginWith{#1}{tab:}{Table~\ref{#1}}{}%
	\IfBeginWith{#1}{appendix:}{Appendix~\ref{#1}}{}%
	\IfBeginWith{#1}{sec:}{Section~\ref{#1}}{}%
}

\newcommand{\etal}{\textit{et al.\ }}
\newcommand{\be}{\begin{align}}
\newcommand{\ee}{\end{align}}
\newcommand{\ket}[1]{\ensuremath{\left| {#1} \right>}}

\newcommand{\Be}{\ensuremath{^9\mathrm{Be}^+\;}}
\newcommand{\BeNoSpace}{\ensuremath{^9\mathrm{Be}^+}}

\newcommand{\CaHp}{\ensuremath{\mathrm{CaH}^+\;}}

\newcommand{\Htwop}{\ensuremath{\mathrm{H}_2^+\;}}

\newcommand{\TtwopNoSpace}{\ensuremath{\mathrm{T}_2^+}}
\newcommand{\HtwopNoSpace}{\ensuremath{\mathrm{H}_2^+}}

\newcommand{\HDp}{\ensuremath{\mathrm{HD}^+\;}}
\newcommand{\HDpNoSpace}{\ensuremath{\mathrm{HD}^+}}

\newcommand{\Htwo}{\ensuremath{\mathrm{H}_2\;}}

\newcommand{\Dtwop}{\ensuremath{\mathrm{D}_2^+\;}}

\newcommand{\HtwopBe}{\ensuremath{\mathrm{H}_2^+ - {^9\mathrm{Be}}^+ \;}}

\newcommand{\He}{\ensuremath{\mathrm{He}\;}}

\newcommand{\Hz}[1]{\SI{#1}{\hertz}}

\newcommand{\MHz}[1]{\SI{#1}{\mega\hertz}}
\newcommand{\GHz}[1]{\SI{#1}{\giga\hertz}}

\newcommand{\ms}[1]{\SI{#1}{\milli\s}}
\newcommand{\us}[1]{\SI{#1}{\micro\s}}

\newcommand{\nm}[1]{\SI{#1}{\nano\meter}}

\usepackage[normalem]{ulem}

\newcommand{\thetitle}{Quantum control of a single \boldmath$\mathrm{H}_2^+$ molecular ion}

\newcommand{\theauthors}{
\author{D. Holzapfel}
\email{dholzapfel@phys.ethz.ch}
\author{F. Schmid}
\author{N. Schwegler}
\affiliation{Department of Physics, ETH Z\"urich, Zurich, Switzerland}%
\affiliation{Quantum Center, ETH Z\"urich, Zurich, Switzerland}%
\author{O. Stadler}
\affiliation{Department of Physics, ETH Z\"urich, Zurich, Switzerland}%
\author{M. Stadler}
\author{A. Ferk}
\author{J. P. Home}
\author{D. Kienzler}
 \email{daniel.kienzler@phys.ethz.ch}
\affiliation{Department of Physics, ETH Z\"urich, Zurich, Switzerland}%
\affiliation{Quantum Center, ETH Z\"urich, Zurich, Switzerland}%
}

\begin{document}
\title{\thetitle}
\theauthors

\begin{abstract}
Science is founded on the benchmarking of theoretical models against experimental measurements, with the challenge that for all but the simplest systems, the calculations required for high precision become extremely challenging. $\mathrm{H}_2^+$ is the simplest stable molecule, and its structure is calculable to high precision. However, studying $\mathrm{H}_2^+$ experimentally presents significant challenges: Standard control methods such as laser cooling are not applicable due to the long lifetimes of its rotational and vibrational states.
Here we solve this issue by combining buffer gas cooling to quench the $\mathrm{H}_2^+$ rovibrational excitation with quantum logic operations between $\mathrm{H}_2^+$ and a co-trapped ‘helper’ ion to control the molecule's hyperfine structure. This enables us to perform pure quantum state preparation, coherent control, and non-destructive readout, which we use to demonstrate high-resolution microwave spectroscopy in the hyperfine structure of $\mathrm{H}_2^+$ with a precision of 2~Hz. Our results pave the way for high precision spectroscopy of $\mathrm{H}_2^+$ in both the microwave and optical domains. Due to the wide applicability of buffer gas cooling, our method provides a general tool for molecular ion species that are hard to control with quantum logic tools alone.

\end{abstract}

\maketitle

The \Htwop ion is the simplest stable molecule and its structure can be calculated ab initio to high precision \cite{17Korobov}. It is therefore an important system to test bound-state three-body quantum electrodynamics, determine fundamental constants, and to search for new physics \cite{2022Schiller, 2023Delaunay, 2024schillerKarr}. However, it is very difficult to study experimentally, and as a result only little high-precision data exists so far. In contrast to the hetero-nuclear isotopologue \HDpNoSpace, the rovibrational levels of \Htwop are extremely long lived \cite{2022Schiller}. This prevents thermalization of the rovibration on experimental time scales and makes it impossible to perform rotational state-preparation by optical pumping as is utilized for \HDp \cite{10Schneider2}. 
Recently two experiments have demonstrated high-precision rotational and vibrational spectroscopy of \Htwop for the first time, using molecular ensembles \cite{2024Doran,2024Schenkel}. Eventually, such molecular ensemble experiments will likely be limited by systematic uncertainties. In contrast, spectroscopy of single trapped ions, one of the most accurate measurement tools in physics, can achieve much lower uncertainties, exemplified by atomic-ion-based optical clocks \cite{16Huntemann,19Brewer}. 
Spectroscopy of a single trapped \Htwop ion is expected to allow for the most precise  measurements of its internal structure \cite{2014Bakalov,14Schiller,16Karr1}. Experiments with single ions are enabled by quantum control techniques such as laser cooling, pure quantum state preparation, coherent control, and non-destructive readout. A direct implementation requires a suitable electronic structure, which \Htwop lacks. Quantum Logic Spectroscopy (QLS) has been developed to implement quantum control for atomic or molecular ion species without such structure \cite{98Wineland1,05Schmidt}. 
In QLS, a well controlled atomic ion is co-trapped with the ion of interest, and quantum logic operations that utilize their mutual Coulomb interaction enable full control over the ion of interest. QLS has recently been implemented for the molecular ions \ensuremath{\mathrm{MgH}^{+}}, \CaHp and \ensuremath{\mathrm{N}_2^+}  \cite{16Wolf,17Chou,2020Sinhal}. 
This progress has enabled several further steps in molecular ion quantum control \cite{2020Chou,2020Lin,2023Collopy,2024Liu}.

We here extend these techniques by combining buffer gas cooling with QLS and by implementing a state preparation scheme that utilizes both microwave and Raman transitions in the hyperfine structure of the molecular ion. This enables us to perform pure quantum state preparation, coherent control and non-destructive state readout of a single \Htwop molecule.
We demonstrate the capabilities of our platform for precision spectroscopy by performing microwave spectroscopy of a hyperfine transition with a statistical uncertainty of \Hz{2}.\\

\section{Experimental implementation} \label{sec:Experimental implementation}
The \Htwop molecule has two identical nuclei which makes it a symmetric, homonuclear molecule. 
The Pauli principle requires that the molecular wavefunction is anti-symmetric under exchange of the two protons. The molecule therefore occurs in two spin isomers: para-\Htwop with total nuclear spin $I=0$ and even rotational quantum numbers $N$, and ortho-\Htwop with $I = 1$ and odd values of $N$.
We here concentrate on the lowest energy rovibrational state of ortho-\Htwop with vibrational quantum number $\nu=0$ and $N=1$.

\begin{figure}
\includegraphics[width=0.48\textwidth]{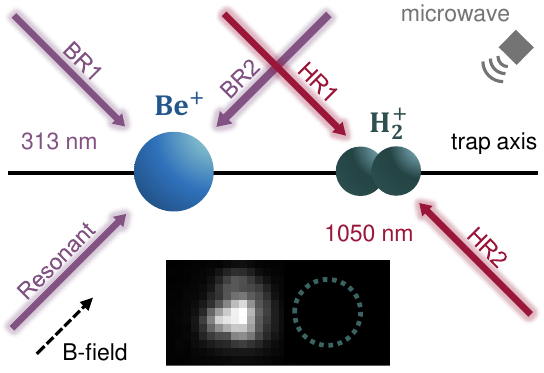}
\caption{\label{fig:setup} \textbf{Sketch of \boldmath$\mathrm{H}_2^+-{^9\mathrm{Be}}^+$ ion pair and beam geometry}. A single \Be and a single \Htwop ion are simultaneously confined in a linear Paul trap and align along the trap axis. A magnetic field (\textbf{B}) with strength $\SI{450}{\micro T}$ defines the quantization axis with an angle of $45^\circ$ with regard to the trap axis. The laser beams addressing \Be at $\SI{313}{nm}$ are shown in purple. The `Resonant' beam is propagating in parallel to the direction of the magnetic field and is used for laser cooling, state preparation and fluorescence detection of \BeNoSpace. The \Be Raman beam pair consists of the `BR1' and `BR2' beams. The \Htwop Raman beam pair at \qty{1050}{nm} (`HR1', `HR2') is shown in red. A microwave drives magnetic dipole transitions in the hyperfine structure of both ions. The inset shows a picture capturing the fluorescence of the \Be ion (left), while the \Htwop ion (right) at a distance of \qty{8.6}{\micro m} remains invisible. The estimated position of the \Htwop ion is indicated by a dotted circle.}
\end{figure}

We co-trap an \Htwop ion and a \Be ion in a linear Paul trap. An ultra-high vacuum chamber with an inner cryogenically cooled chamber houses the ion trap. 
\autoref{fig:setup} shows the geometry of the laser beams used to address \Be and \HtwopNoSpace. A beam that is resonant with the D2 transition (wavelength \nm{313}) is used for Doppler cooling, state preparation and fluorescence detection of the \Be ion. A pair of Raman beams (also \nm{313}) is used to implement coherent control on the two-level system $\ket \downarrow = \ket{F = 2, m_F = 2}$ and $\ket \uparrow = \ket{F = 1, m_F = 1}$ in the hyperfine structure of the \Be electronic ground state.
Two infrared laser beams (wavelength $\nm{1050}$) directed at the \Htwop ion are used to implement stimulated Raman transitions within its hyperfine structure. 
Both Raman beam pairs can couple the internal degree of freedom of the respective ion to a shared motional mode of the ion pair by driving the corresponding motional sideband. This implements the required operations for motional ground-state cooling and QLS.
A microwave antenna located inside the vacuum chamber is used to drive transitions in the hyperfine structure of either ion in a frequency range of $1.2-1.4$~GHz. Due to the long wavelength of the microwaves and the resulting small Lamb-Dicke parameter no coupling to the ions' motion is possible.
For more details on the apparatus and experimental techniques see the \hyperref[Methods:A]{Appendix A} and Schwegler \etal \cite{2023Schwegler}.

We produce \Htwop ions using electron-impact ionization of \Htwo from the background gas. This results in a wide distribution of vibrational and rotational states \cite{93Weijun}.
The symmetric, homonuclear nature of \Htwop causes its rovibrational levels to be extremely long-lived as rovibrational dipole transitions are strongly suppressed \cite{1983Posen,13Pilon,75Bishop,2023Korobov}. While this is a desirable feature for high-precision spectroscopy, it prevents the rotation and vibration from thermalizing on experimental timescales. To solve this issue and prepare our rovibrational target state ($\nu=0,N=1$), we utilize helium buffer gas cooling of a single \Htwop \cite{14Hansen,17Schiller}. The vacuum chamber is equipped with a leak valve to provide helium gas. We have leaked helium gas into the chamber once and have observed helium in the chamber since. The helium background gas is cooled by collisions with the inner cryogenic chamber. 
We cannot measure the \He pressure in the cold chamber with a pressure sensor, but collisions of the trapped ions with helium atoms can cause the \Be and \Htwop to switch places in the trap. The ion-reorder rate thus is a proxy for the pressure. We measure the ion-reorder rate by recording the \Be ion position on a camera. We observe a strong asymmetry in the reorder rate depending on the initial order of the ion pair. We attribute this to a tilted potential well, likely caused by electrostatic charging of the trap due to the electron beam, causing a energetically preferred ion pair order. At cryostat temperatures of \qtylist{5;8;10}{\kelvin} we measure reorder rates of \qtylist{0.004(1); 0.05(1);0.26(7)}{s^{-1}}, respectively (averaged over the ion pair order, errors are standard errors of the Poisson distributions). Detailed modeling of the collision dynamics would be required to derive the helium gas pressure from the reorder rate \cite{2019Hankin}, especially due to the low kinetic energy of the helium buffer gas, and is not performed here. To conserve liquid helium required for the cooling of the cryostat, all data in this study was taken at a cryostat temperature of \qty{10}{\kelvin}, but lower temperatures could be used to lower the collision rate if collisions limit the accuracy of \Htwop spectroscopy \cite{2019Hankin}.

The \nm{313} light used for \Be control dissociates \Htwop with rates strongly dependent on the vibrational state. From \nm{313} laser intensities estimated at the \Htwop ion's location (see \hyperref[Methods:A]{Appendix A}), its vibrationally excited states of $\nu \ge 2$ are expected to have a lifetime of less than one second, while $\nu = 1$ and $\nu = 0$ have lifetimes of a few minutes and several days, respectively  \cite{dunn1968a,PCKarr2024}. Thus, \Htwop with $\nu \ge 2$ is dissociated before buffer gas cooling has a significant impact \cite{17Schiller}.

Based on the time it takes to observe the first quantum logic signal of the $\nu=0, N=1$ hyperfine structure (see below) after \Htwop is loaded, we estimate an approximate duration of $\mathcal{O}\left(\qty{10}{min}\right)$ for the rovibrational state to be cooled to the ground state. In contrast, the unperturbed lifetime of the $\nu = 1$ state is 22 days \cite{1983Posen}.
In some instances we lose \Htwop in $<\qty{10}{min}$. We attribute this loss to dissociation of the $\nu = 1$ state due to \nm{313} laser light before buffer gas cooling to $\nu=0$ could occur. 
For approximately half of the \Htwop molecules loaded, we do not observe a quantum logic signal after ten minutes. These molecules are likely to be para-\HtwopNoSpace, for which our quantum logic scheme does not provide a signal.
Since electron impact ionization of \Htwop is not isomer-selective, we expect para- and ortho-\Htwop to be loaded randomly. We do not know the ortho-para ratio of the \Htwo background gas, which is likely composed of \Htwo molecules entering the cryogenic chamber from the room-temperature part of the vacuum chamber, \Htwo thermalized with the cold chamber, and hot \Htwo emitted from the electron gun filament.

\section{\boldmath$\mathrm{H}_2^+$ hyperfine structure} 
\begin{figure}[h!]
\includegraphics[width=0.48\textwidth]{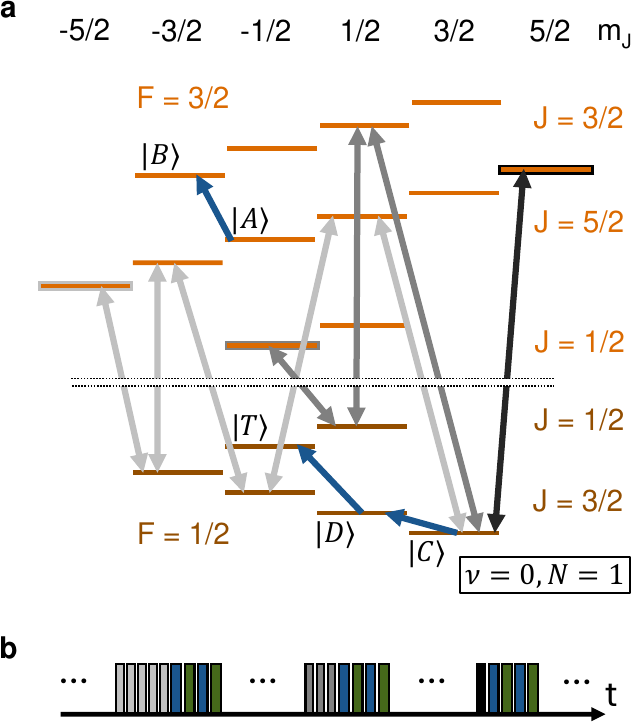}
\caption{\label{fig:Htwop_Structure} \textbf{Relevant energy level structure of ortho-\boldmath$\mathrm{H}_2^+$ and QLS scheme.} \textbf{a}: Hyperfine and Zeeman structure of ortho-\Htwop in its vibrational ground state $\nu=0$ and lowest rotational state $N=1$. Each of the 18 hyperfine states is labeled with its dominant $\ket{F,J,m_J}$ component. States in the $\ket{F=1/2}$ manifold are indicated in brown, while orange indicates states in the $\ket{F=3/2}$ manifold. The energy splittings between the states are not shown to scale. The energy difference between the $F=1/2$ and $F=3/2$ manifolds is $1.3-\SI{1.4}{GHz}$ while states within an $\ket{F,J}$ manifold are split by up to $\qty{13}{MHz}$ at our magnetic bias field of \qty{450}{\micro\tesla}. To observe a first quantum logic spectroscopy signal, we probe with a Raman blue sideband transition (blue arrow) between states $\ket{A}$ and $\ket{B}$. For pure state preparation in the hyperfine manifold, Raman blue sideband transitions (blue arrows) are used to implement irreversible population transfer from $\ket C$ via $\ket D$ into the target state \ket{T} using QLS operations. Different microwave transition chains connect each state of the hyperfine manifold to the Raman transitions. Three examples of such chains for transferring the population from the states \ket{3/2,5/2,-5/2} (light gray arrows), \ket{3/2,1/2,-1/2} (dark gray arrows), and \ket{3/2,5/2,5/2} (black arrow) to \ket{C} are shown. \textbf{b}: Exemplary state preparation pulse sequences. Each group of pulses pumps the population of one specific state (first \ket{3/2,5/2,-5/2}, second \ket{3/2,1/2,-1/2}, third \ket{3/2,5/2,5/2}) to the target state \ket{T}. Microwaves transfer the population from the starting state to \ket{C}. Then, Raman blue sideband transitions on \Htwop (blue) and pulses for implementing dissipation on \Be (green) pump the population from \ket{C} via \ket{D} to \ket{T}. Relative pulse times are not shown to scale.}
\end{figure}

After the rovibrational state $\nu=0,N=1$ is prepared, its hyperfine structure is in a mixed state. \hyperref[fig:Htwop_Structure]{Figure 2.a} shows the 18 hyperfine states. We use the angular momentum coupling scheme
\begin{equation}
    \mathbf{F} = \mathbf{S}_e + \mathbf{I} \qquad \text{and} \qquad \mathbf{J} = \mathbf{L} + \mathbf{F},
\end{equation}
where $\mathbf{S}_e$ is the electron spin, $\mathbf{I}$ is the total nuclear spin, and $\mathbf{L}$ is the total orbital angular momentum \cite{karr2008}. Since we work with \Htwop in its electronic ground state, $\mathbf{L}$ is equal to the rotational angular momentum $\mathbf{N}$. At a finite magnetic field, neither $F$ nor $J$ are good quantum numbers. However, at our magnetic bias field of \qty{450}{\micro\tesla}, each hyperfine state has one strongly dominant $\ket{F, J, m_J}$ component, 
and we use this component to label the state. 
The frequency difference between the $F=1/2$ and $F=3/2$ manifolds is $1.3-\SI{1.4}{GHz}$ while states within an $\ket{F,J}$ manifold are split by up to $\qty{13}{MHz}$.\\

\section{Quantum logic spectroscopy\\ of \boldmath$\mathrm{H}_2^+$}\label{sec:firstdetect}
To observe a first signal from the hyperfine structure of \HtwopNoSpace, we probe a single hyperfine transition of \Htwop with QLS and observe the signal over time.

For all QLS operations we use the \Be hyperfine states $\ket{\downarrow}$ and $\ket{\uparrow}$ and the ion pair's axial out-of-phase mode of motion (with Fock states \ket{n}).
In this initial experiment we choose to probe the \Htwop transition $\ket{A}\equiv\ket{3/2,5/2,-1/2}\leftrightarrow\ket{B}\equiv\ket{3/2,3/2,-3/2}$, as Raman transition driven by the \nm{1050} laser beams (see \hyperref[fig:Htwop_Structure]{Figure 2.a})). This transition features one of the strongest matrix elements of the $N=1$ hyperfine manifold and has a frequency of \MHz{5.6}.
\begin{figure}[h!]
\includegraphics[width=0.48\textwidth]{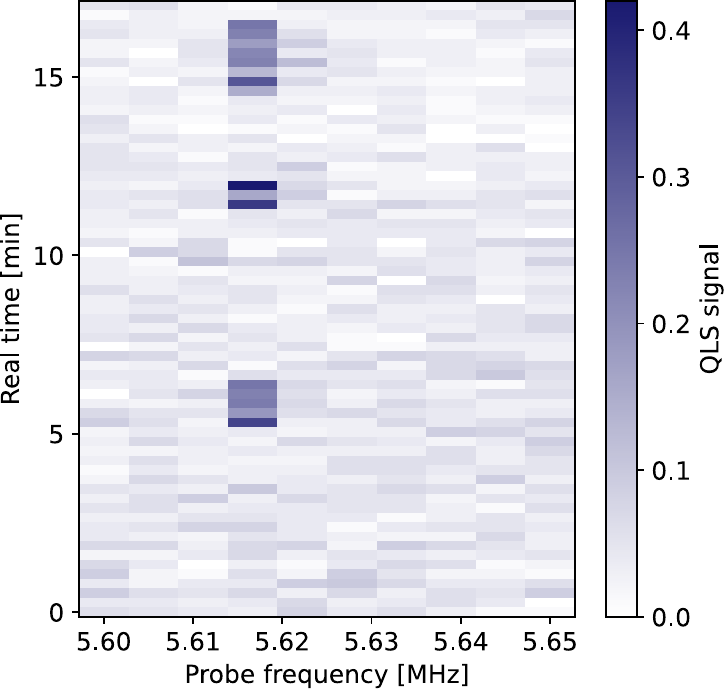}
\caption{\label{fig:qls_detect} \textbf{Quantum logic spectroscopy of ortho-\boldmath$\mathrm{H}_2^+$ hyperfine transition.} Observation of the QLS signal on the $\ket{A}\equiv\ket{3/2,5/2,-1/2}\leftrightarrow\ket{B}\equiv\ket{3/2,3/2,-3/2}$ hyperfine transition over a time span of $\approx \SI{16}{min}$, by scanning the \Htwop blue sideband frequency over the expected transition frequency. The frequency scans are performed left to right, and the data consists of 65 such frequency scans performed sequentially after each other, bottom to top, while real time is passing. Each data point consists of 100 iterations of the experiment. At random time intervals, a QLS signal appears at the transition frequency. We attribute the random appearance and disappearance of the signal to collisions with the He gas changing the hyperfine state, spontaneously populating different hyperfine levels.}
\end{figure}

We read out the \ket{A} state population by first performing ground-state cooling of the out-of-phase mode and state preparation on \BeNoSpace, initializing the ion pair and its motion in \ket{\Psi,\downarrow, n=0}, where \ket{\Psi} is the hyperfine state of \HtwopNoSpace. In the case that $\ket \Psi = \ket{A}$, a blue-sideband $\pi$-pulse on the \Htwop $\ket{A}\leftrightarrow\ket{B}$ transition changes the full system's state from \ket{A,\downarrow,0} to \ket{B,\downarrow,1}, exciting the shared motion. For $\ket \Psi \neq \ket{A}$ the shared motion is not excited. The motional state can be detected using a subsequent red sideband $\pi$-pulse and state-dependent fluorescence detection of \BeNoSpace. We call the resulting measurement probability to find \Be in the $\ket{\uparrow}$ state the `QLS signal'. It is proportional to the transition probability of the \Htwop $\ket{A}\leftrightarrow\ket{B}$ transition and thus the \ket{A} state population before the QLS readout sequence. This basic sequence would provide only a signal once, as it results in transferring the population from \ket{A} to \ket{B} as part of the readout sequence. To enable repetition of the experiment we perform a reset operation after the \Be red sideband $\pi$-pulse by driving a $\pi$-pulse on the bare ``carrier'' $\ket{A}\leftrightarrow\ket{B}$ transition, without coupling to the motion. This swaps the population of the states \ket{A} and \ket{B}, and thus returns the population of \ket{B} to \ket{A}, enabling a repetition of the sequence. 
We choose this simple method in this initial experiment, as it is sufficient to observe a first QLS signal while being less complex and calibration reliant than sideband-based reset schemes. 
To see a QLS signal from the transition and observe it over time, we probe the transition with identical settings for 100 repetitions of the simplified QLS experiment. We scan the frequency around the expected theoretical value and repeat this frequency scan over time. The transition frequency is scanned simultaneously for the blue sideband and the carrier transition. The duration to acquire a single data point is $\SI{1.6}{s}$, and, thus, $\SI{16}{s}$ for one frequency scan. The resulting data is shown in \hyperref[fig:qls_detect]{Figure 3}.

The signal of the $\ket{A}\leftrightarrow\ket{B}$ transition appears and disappears over time, with an approximate dwell-time of a few seconds. We attribute this behavior to collisions with the He gas changing the hyperfine state, spontaneously populating different hyperfine levels (see \hyperref[sec:coherent control]{Section V}). Since the probe sequence cycles population between two out of 18 hyperfine levels, we expect the signal to appear on average 11\% of the probe duration. The data presented here is selected to show multiple re-appearances over time and is not representative. Some data sets not shown here do not display any signal over the duration probed here.

Fundamentally, it would be possible to perform spectroscopy already with this spontaneously occurring signal. However, to improve signal strength, in the following we implement an active, deterministic state-preparation of a single hyperfine state.

\section{Pure quantum state preparation\\ of \boldmath$\mathrm{H}_2^+$}\label{sec:state prep}
To prepare \Htwop deterministically in a pure quantum state, we employ a state preparation sequence that uses quantum logic operations and collects the population from all hyperfine states into a target state $\ket{T} = \ket{F = 1/2, J = 1/2, m_J = -1/2}$. 
For the QLS readout we choose the \Htwop transition $\ket{T}\leftrightarrow\ket{D}=\ket{1/2,3/2,1/2}$, which can be driven as a stimulated Raman transition by the \nm{1050} beams and has a frequency of \MHz{20.4}.

We read out the \ket{T} state population by first performing ground-state cooling of the out-of-phase mode and state preparation on \BeNoSpace, initializing the ion pair and its motion in \ket{\Psi,\downarrow, n=0}, where \ket{\Psi} is the hyperfine state of \HtwopNoSpace. In the case that $\ket \Psi = \ket{T}$, a red-sideband $\pi$-pulse on the \Htwop $\ket{T}\leftrightarrow\ket{D}$ transition changes the full system's state from \ket{T,\downarrow,0} to \ket{D,\downarrow,1}, exciting the shared motion. For $\ket \Psi \neq \ket{T}$ the shared motion is not excited. The motional state can be detected using a subsequent red sideband $\pi$-pulse and state-dependent fluorescence detection of \BeNoSpace. We call the resulting measurement probability to find \Be in the $\ket{\uparrow}$ state the `QLS signal'. It is proportional to the transition probability of the \Htwop $\ket{T}\leftrightarrow\ket{D}$ transition and thus the \ket{T} state population before the QLS readout sequence.

Within the two-level subsystem $\{\ket{D},\ket{T}\}$ of \HtwopNoSpace, the state \ket{T} can be prepared with a similar pulse sequence, replacing the red sideband with a blue sideband on the $\ket{D}\leftrightarrow\ket{T}$ transition and performing a repump operation on \Be instead of the fluorescence detection. This pumps the population from the two-level system $\{\ket{D},\ket{T}\}$ into \ket{T}. 
Pure quantum state preparation can be achieved by relying exclusively on QLS pumping operations, implementing them on the full hyperfine structure as was demonstrated for \CaHp \cite{17Chou}.
This however is hard to achieve for the \Htwop molecule as matrix elements for Raman transitions connecting the $F = 1/2$ and $F = 3/2$ hyperfine manifolds are at least one order of magnitude weaker than transitions within the $F = 1/2$ manifold. This is a consequence of the Raman transitions only coupling to the orbital angular momentum, not the spins and the $F$-mixing being weak at our magnetic field.

We solve this issue by implementing stimulated Raman transitions and their motional sideband operations only on two hyperfine transitions ($\ket{C} = \ket{1/2,3/2,3/2} \leftrightarrow\ket{D}$ and $\ket{D}\leftrightarrow\ket{T}$, see \hyperref[fig:Htwop_Structure]{Figure 2.a}). This two-link chain collects population from the states \ket{C} and \ket{D} into \ket{T}. To extend this to all hyperfine states we use microwaves driving magnetic dipole transitions in the hyperfine structure. With sequences of microwave $\pi$-pulses we can efficiently connect all levels, but due to the microwave's long wavelength and corresponding small Lamb-Dicke parameter, they cannot be used to implement the sideband operations required for quantum logic.
We therefor alternate microwave operations transferring population from other levels into state \ket{C} and Raman sideband operations that move the population of levels \ket{C} and \ket{D} into \ket{T}. The full state preparation sequence consists of 16 repetitions of the $\ket{C}\rightarrow\ket{D}\rightarrow\ket{T}$ chain, with a unique microwave pulse sequence before each repetition. Each of these 15 unique microwave pulse sequences transfers population from a different state to \ket{C}. Three examples for such microwave pulse sequences are shown in different shades of gray in \hyperref[fig:Htwop_Structure]{Figure 2.a}. An exemplary sequence that combines the microwave pulses with QLS operations is shown in \hyperref[fig:Htwop_Structure]{Figure 2.b}.
We compute the optimal sequence to transfer all states to \ket{T} with a shortest-path algorithm \cite{23Stadler}. This takes into account that microwave pulses can only perform reversible operations and thus can only be used in the state preparation to transfer population to previously emptied states.

\begin{figure*}
\includegraphics[width=0.98\textwidth]{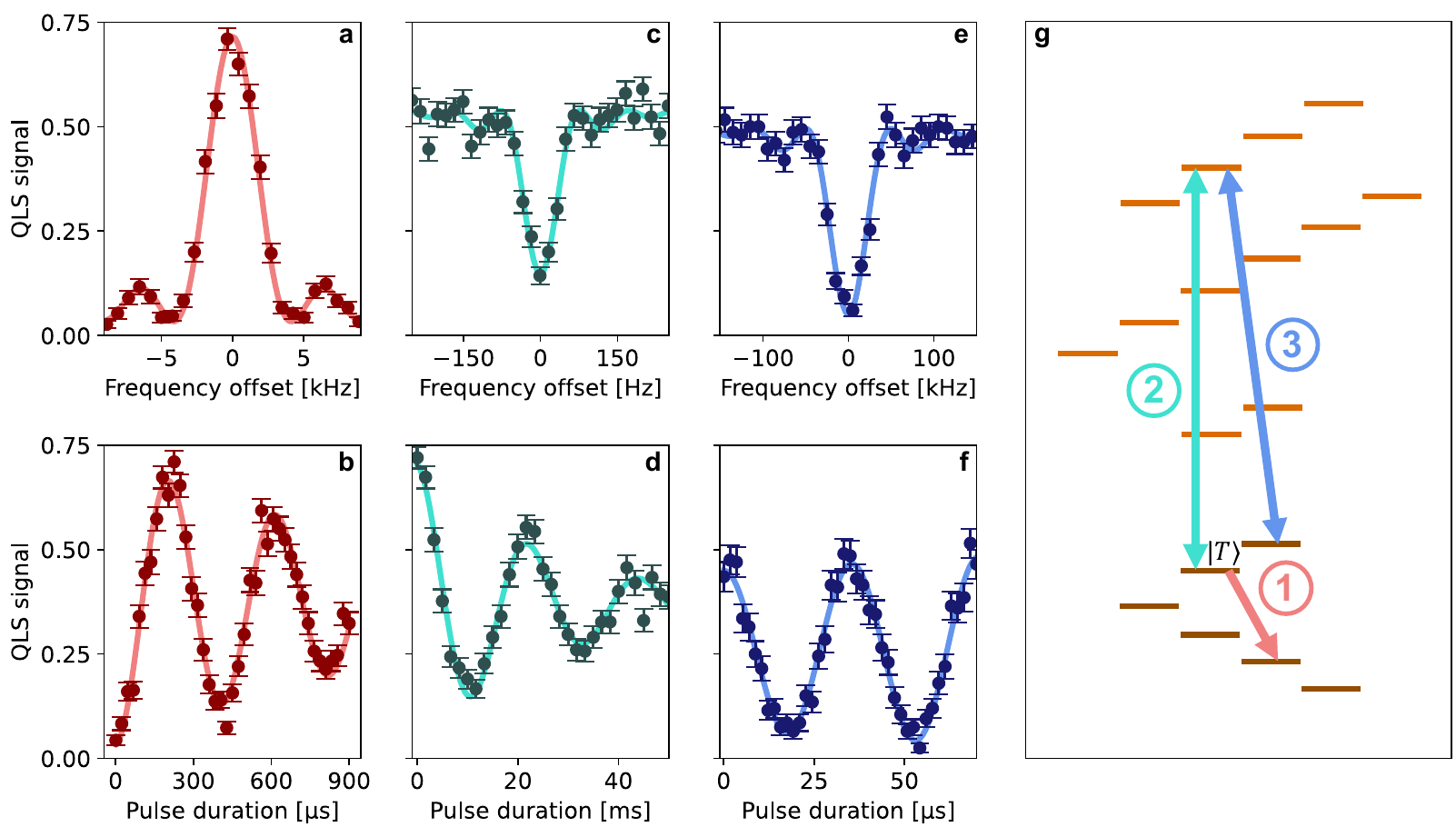}
\caption{\label{fig:alldata} \textbf{Coherent manipulation of pure hyperfine states of ortho-\boldmath$\mathrm{H}_2^+$}. In subfigures a-e, each data point corresponds to 300 repetitions of the respective sequence, and in subfigure f one data point is 200 repetitions. The QLS signal is the probability to find \Be in the $\ket{\uparrow}$ state after the QLS readout and is proportional to the \ket{T} population before QLS readout. Error bars indicate the statistical uncertainty (Standard Error of the Mean, SEM). Solid lines are fits to the data (sinc$^2(\pi/2\sqrt{1+4(\delta \, t)^2})$ with the frequency offset $\delta$ and the $\pi$-pulse duration $t$ for a,c,e \cite{ramsey1956}; cosine function with exponential damping for b,d,f) \textbf{a}, \textbf{b}: Raman red sideband frequency spectrum (with Raman pulse duration \us{210}) and Rabi oscillations of transition 1. The $\pi$-pulse contrast extracted from the fit to the Rabi oscillation data is $0.665(8)$. \textbf{c}, \textbf{d}: Microwave frequency spectrum (microwave pulse duration \ms{11}) and Rabi oscillations of the magnetic field insensitive transition 2 using QLS readout on transition 1. \textbf{e}, \textbf{f}: Microwave frequency spectrum  (microwave pulse duration \us{17.4}) and Rabi oscillations of transition 3 using QLS readout on transition 1 and $\pi$-pulses on transition 2 to connect transition 3 to transition 1. The fits result in center frequencies of $\SI{20.42638(4)}{MHz}$ for transition 1, $\SI{1392.081049(2)}{MHz}$ for transition 2 and $\SI{1391.3083(7)}{MHz}$ for transition 3. \textbf{g}: Schematic of the hyperfine structure of the rovibrational ground state of ortho-\Htwop indicating transitions 1, 2 and 3.}
\end{figure*}

In principle, only one Raman sideband pumping transition combined with microwave sequences is sufficient for the protocol. We however observe lower state preparation contrast using only one transition. This is expected as the ground-state cooling of the ions' motion is imperfect, leading to a small leakage from \ket{T} to \ket{D}. The issue is amplified by the reversible nature of the microwave operations, spreading the leaked population over the full hyperfine structure. Using a second Raman transition suppresses this effect from a linear dependence on the ground-state cooling infidelity for a single Raman transition to a quadratic dependence for two Raman transitions, `shielding' the population in \ket{T}. For the $\ket{A}\leftrightarrow\ket{B}$ transition no strong connecting second Raman transition exists, which is why we implement this scheme with the states \ket{C}, \ket{D}, \ket{T} instead. As demonstrated, other hyperfine states can be prepared using microwaves to transfer from \ket{T}, enabling full control over the hyperfine structure.

\section{Coherent control of \boldmath$\mathrm{H}_2^+$}\label{sec:coherent control}
To test the preparation of \ket{T} we perform the state preparation sequence followed by the quantum logic readout of the \ket{T} state population on the $\ket{T}\leftrightarrow \ket{D}$ transition (transition 1 in \hyperref[fig:alldata]{Figure 4.g}), as described above. We perform a frequency and pulse duration scan of the \Htwop red sideband operation used in the QLS readout, shown in \hyperref[fig:alldata]{Figures 4.a and b}, demonstrating Rabi oscillations between the levels \ket{T} and \ket{D}. From the Rabi oscillation data, we extract a $\pi$-pulse contrast of 0.665(8), which corresponds to the combined state-preparation and readout fidelity. Additionally performing a postselection of the data based on a heralding detection, we achieve an improved contrast of 0.74(1) (see \hyperref[Methods:Herald]{Appendix E} for details and postselected data in \hyperref[fig:heralded]{Figure 8}). The decay of the Rabi oscillations is consistent with averages of Debye-Waller couplings to the non-ground state cooled radial modes of motion \cite{98Wineland1}. The 7.6\% improvement of the signal contrast using the heralded detection suggests that the \Htwop hyperfine state preparation has a fidelity of above $0.9$, and that imperfect QLS readout is causing $\approx 26\%$ signal contrast reduction. Known contributions to the imperfect QLS readout are effects due to the residual motion of the ion pair such as imperfect ground-state cooling of the axial out-of-phase mode of motion and a residual Debye-Waller effect from the not ground-state cooled radial motional mode and the not perfectly ground-state cooled axial in-phase motional mode. Implementing additional cooling pulses could improve the signal contrast.

To demonstrate the spectroscopy capability of our method, we prepare the state \ket{T} and 
perform microwave spectroscopy on the $\ket{T}\leftrightarrow\ket{3/2,3/2,-1/2}$ transition (transition 2 in \hyperref[fig:alldata]{Figure 4.g}). The data is shown in \hyperref[fig:alldata]{Figure 4.c and d}. The readout is again performed with QLS, probing the population of \ket{T} on transition 1. At our chosen magnetic field (\qty{450}{\micro\tesla}), the frequency of transition 2 has only a weak dependence on the magnetic field (\qty{-31}{Hz/\micro\tesla}) \cite{karr2008}, which allows us to coherently drive the transition with a \ms{11} $\pi$-pulse, resulting in a Fourier-limited full width at half-maximum linewidth of \Hz{71}. By averaging for half an hour we achieve a statistical uncertainty of \Hz{2} which corresponds to a relative uncertainty of $1 \times 10^{-9}$.
The decay of the Rabi oscillations (\hyperref[fig:alldata]{Figure 4.d}) is consistent with the combined heating of the in-phase axial and radial modes of motion during the microwave probe pulse.

To prepare other hyperfine states and perform spectroscopy of hyperfine transitions which do not include the initial state \ket{T}, it is possible to add a sequence of microwave $\pi$-pulses after the preparation of \ket{T}.
We demonstrate this by performing spectroscopy on the $\ket{3/2,3/2,-1/2}\leftrightarrow\ket{1/2,1/2,1/2}$ transition (transition 3 in \hyperref[fig:alldata]{Figure 4.g}). We first prepare \ket{T}, followed by a microwave $\pi$-pulse on transition 2, transferring the population to \ket{3/2,3/2,-1/2}. We then drive transition 3 and return the residual population of \ket{3/2,3/2,-1/2} to \ket{T} with another transition 2 microwave $\pi$-pulse followed by QLS readout of \ket{T}. The data is shown as a function of the duration and frequency of the pulse on transition 3 in \hyperref[fig:alldata]{Figure 4.e and f}. 

\begin{figure}
\includegraphics[width=0.48\textwidth]{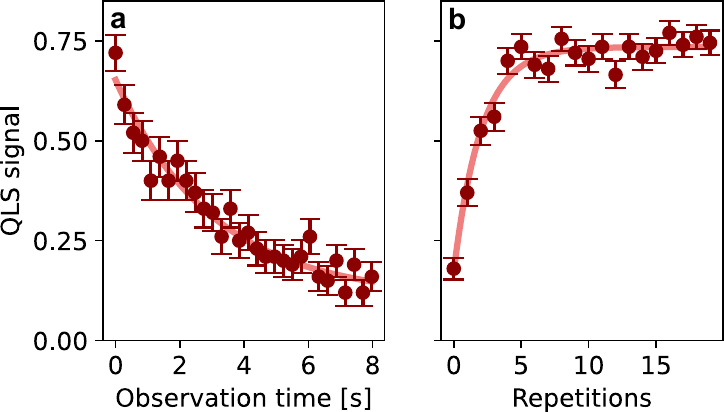}
\caption{\label{fig:testpump_depump} \textbf{Ortho-\boldmath$\mathrm{H}_2^+$ hyperfine state loss and recovery.} The QLS signal is the probability to find \Be in the $\ket{\uparrow}$ state after the QLS readout and is proportional to the \ket{T} population before QLS readout. \textbf{a}: Loss of the \ket{T} state population over time. The \Htwop ion is initially prepared in the state \ket{T}. After this the state is repeatedly probed with QLS readout for up to 8 seconds. The full procedure is repeated 100 times and each data point corresponds to the average at a given duration after the state preparation, with error bars indicating the statistical uncertainty (SEM). We extract a $1/e$ lifetime of \qty{3.1(6)}{\s} from an exponential fit to the data (solid line). 
\textbf{b}: Number of state preparation sequence repetitions required to prepare the \Htwop molecule in state \ket{T}. Initially, a wait time of \qty{10}{\s} allows collisions with background gas to randomize the hyperfine state of the \Htwop molecule. Subsequently, we apply the state preparation sequence a given number of times and read out the population of \ket{T}. Fitting an exponential function (solid line), we find the 1/e value to be \qty{2.1(2)}{repetitions}. Each data point is the average of 200 repetitions of the sequence, with error bars indicating the statistical uncertainty (SEM).
}
\end{figure}

To investigate hyperfine state changes over time we prepare state \ket{T} and repeatedly probe it for a duration of up to \qty{8}{\s} without re-preparation. The results are shown in \hyperref[fig:testpump_depump]{Figure 5.a}. The data follows an exponential decay, and we extract a $1/e$ lifetime of \qty{3.1(6)}{\s} from the fit. All data presented in this article was taken at a cryostat temperature of \qty{10}{\kelvin}. We have however observed a variation in the $1/e$ lifetime for different cryostat temperatures and different durations after ion loading. This suggests a dependence of the hyperfine state lifetime on the helium pressure. We suspect that cyro-adsorbed helium is released during the electron-impact ionization of \Htwo and then slowly cryogenically pumped without reaching a steady state for several hours. 
As a complimentary indication of such behavior we have observed that the ion reorder rate decreases with lower cryostat temperatures, which is consistent with a lower pressure.
We suspect that the hyperfine state changes are caused either by the collisions directly or by the collisions driving the molecule out of the RF null of the Paul trap and thereby enhancing off-resonantly driven hyperfine transitions by the RF field \cite{17Vera1}. 

Similarly, we can examine the performance of the \Htwop hyperfine state preparation. For this we wait for a fixed duration of $\SI{10}{s}$ to ensure a random initial state. Thereafter, we run the state preparation sequence a given number of times and probe the achieved population of the state \ket{T} (see \hyperref[fig:testpump_depump]{Figure 5.b}). The QLS signal increases for increasing numbers of state preparation sequence repetitions and saturates after approximately five repetitions. From an exponential fit, we extract a 1/e value of the state recovery of \qty{2.1(2)} repetitions. The contrast gain per state preparation cycle is likely limited by the many imperfect operations employed in the pumping sequence. We suspect the strongest impact to be from the microwave operations in the state preparation. Performing microwave spectroscopy of several hyperfine transitions of \Be using the same microwave input power, we observe a disagreement with theoretical calculations of the matrix elements (leaving the polarization as a free parameter). This suggests that the microwave intensity at the ions' location is frequency dependent. Since the pulse duration of all microwave pulses used in the \Htwop state-preparation are derived from only three measurements (see \hyperref[Methods:Pumping]{Appendix D}), we expect that the executed pulses do not implement $\pi$-pulses for all microwave transitions, making the state transfer imperfect. This could be improved by directly calibrating each transition or by calibration of the frequency-dependent microwave intensity at the ion location by other means.

\section{Outlook}
Without further improvements of our method, it is possible to perform high-precision hyperfine spectroscopy of the $\nu=0, N=1$ state. The hyperfine structure of \Htwop has been measured in ensemble experiments \cite{1968Richardson,1969Jefferts,1992Fu,2004Osterwalder}, with the most accurate published measurement by Jefferts with a relative uncertainty of $\ge1.2\times10^{-6}$ \cite{1969Jefferts}. Additionally, unpublished results by Menasian and Dehmelt with a relative uncertainty of $1.5\times10^{-7}$ exits, however of only few, selected transitions \cite{1973Menasian,ThMenasian}. We anticipate that a measurement with our approach would allow to achieve a better accuracy over a wider range of transitions. These measurements would allow a more complete comparison to theory \cite{haidar2022}, and in the future may provide the starting point for a matter-anti-matter comparisons to test CPT \cite{18Myers}.
Beyond this, our work paves the way for a broader range of future high-precision studies of the \Htwop structure. The state-preparation scheme can be used as a starting point to implement single-ion laser spectroscopy of the \Htwop rovibrational structure. For selected transitions this is projected to reach a relative uncertainty of $10^{-16} - 10^{-17}$, three to four orders of magnitude beyond the state of the art, which provides the experimental sensitivity to improve the uncertainty of several fundamental constants such as the proton-electron mass ratio significantly \cite{2014Bakalov,14Schiller,16Karr1}. The uncertainty for vibrational, rotational, and hyperfine spectroscopy will likely outperform the current uncertainty of the theory. Therefore, theory improvements and spectroscopy schemes that suppress the theoretical uncertainty will be of high importance to harness the potential of this platform \cite{2024Schiller}. 
The techniques we have demonstrated in this study are not limited to \Htwop and can be implemented for other molecular ion species. The combination of buffer gas cooling with quantum logic techniques could enable pure quantum state preparation for a wide range of molecular ions which cannot easily be controlled by QLS alone. As demonstrated here, it provides a technique to quench rovibrational excitation in homo-nuclear molecules and can be applied to the other homonuclear molecules, for instance the isotopologues of the hydrogen molecular ion \Dtwop and \TtwopNoSpace. Spectroscopy of the \Dtwop hyperfine structure would allow to perform a measurement of the electric quadrupole moment of the deuteron \cite{2021Danev}.
Buffer gas cooling can also be utilized to cool the rotation of polar molecules below the temperature of the black body radiation spectrum, which could broaden the applicability of established QLS techniques to complex polyatomic molecules.\\

During the preparation of this manuscript, we became aware of a study demonstrating non-destructive control of an HD$^+$ molecular ion in a Penning trap \cite{25konig}.

\section*{Acknowledgments}
The authors thank J. C. J. Koelemeij and J.-Ph. Karr for helpful discussions and the Segtrap and Penning teams of the ETHZ TIQI group for sharing laser light. This work was supported by Swiss National Science Foundation Grant No. 179909 and 212641, as well as ETH Research Grant No. ETH-52 19-2.

\section*{Appendix A: Apparatus and ion control}\label{sec:Methods}
\label{Methods:A}

Our apparatus, ion loading procedure, \Htwop lifetimes and motional ground-state cooling are described in detail in \cite{2023Schwegler}. Some key features are summarized in the following. Additionally, we will point out significant changes and additions to the apparatus.

The trap is a micro-fabricated monolithic linear Paul trap with an electrode-ion distance of $\SI{300}{\micro m}$. It is operated at a frequency of \SI{78.5}{MHz}. The  motional frequencies of the axial in- and out-of-phase modes of the \HtwopBe ion crystal are \SI{1.3}{MHz} and \SI{3.4}{MHz}, respectively. Due to the large mass mismatch of the two ions, their radial motion is only very weakly coupled \cite{12Wubbena}. We can, therefore, allocate each of the radial motional modes to primarily one of the two ions. The two out-of-phase radial modes where primarily the \Be ion moves have frequencies \SI{1.7}{MHz} and \SI{1.9}{MHz}, the modes dominated by \Htwop are the in-phase modes at 
\SI{9.6}{MHz} and \SI{9.8}{MHz}.

The usual control sequence to prepare both the motional and the internal states of the two ions is: Doppler cooling of \Be and initialization of ion order, optical dipole force gradient (ODF) assisted Doppler cooling of the \Htwop dominated radial motional modes, ground-state cooling of the axial motional modes, \Be state preparation to $\ket \downarrow = \ket{F = 2, m_F = 2}$, \Htwop hyperfine state preparation (\hyperref[Methods:Pumping]{Appendix D}) to $\ket T$, QLS detection to optionally herald (\hyperref[Methods:Herald]{Appendix E}) the preparation of state $\ket T$, re-cooling of all motional modes (including Doppler cooling, ODF assisted Doppler cooling and axial ground-state cooling), \Be state preparation, \Htwop state reset after heralding detection, experiment-specific pulses, QLS detection. Details on each step can be found below or in \cite{2023Schwegler}.

The laser beam geometry is shown in \autoref{fig:setup}. The \nm{313} beams manipulating \Be are focused to a beam waist radius of $\approx \SI{10}{\micro m}$ for the Resonant beam and $\approx \SI{16}{\micro m}$ for the two Raman beams BR1 and BR2. The Raman beams and have powers of $\approx \SI{100}{\micro W}$ each. The strongest resonant beam component implements far-detuned Doppler cooling and optical pumping and has a power of $\approx \SI{750}{\micro W}$. This Resonant beam component dominates the photo-dissociation of the \Htwop in exited vibrational states (see \hyperref[sec:Experimental implementation]{Section I}). The Raman beams are detuned from the D2 transition by \GHz{42}. The \nm{1050} beams (HR1 and HR2) used for manipulating \Htwop have a power of $\approx \qty{400}{mW}$ each and are focused to a beam waist radius of $\approx \SI{5}{\micro m}$ at the ion's position. The beam alignment procedure is described in \hyperref[Methods:beam alignment]{Appendix B}. HR1 has $\pi$ polarization while the polarization of HR2 is set approximately half way between $\pi$ and $\left(\sigma_+ + \sigma_- \right)$. Driving the stimulated Raman transitions in the \Htwop hyperfine structure requires polarization $\left(\sigma_+ + \sigma_- \right)$ while the ODF gradient assisted Doppler cooling (\hyperref[Methods:excooling]{Appendix C}) requires $\pi$ polarization. The setting used here is a compromise to enable both operations within the same sequence. 

\section*{Appendix B: Alignment of infrared laser beams}\label{Methods:beam alignment} 
\begin{figure}
\includegraphics[width=0.48\textwidth]{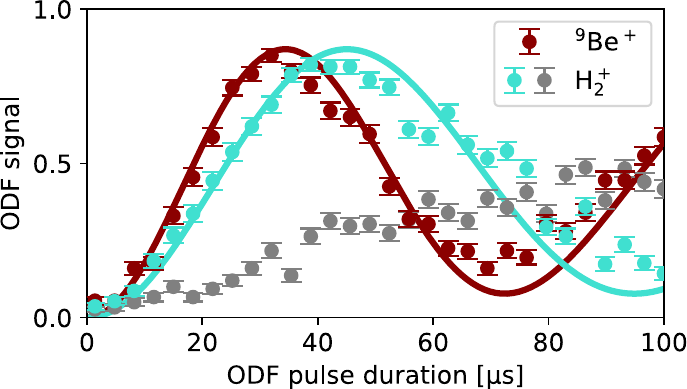}
\caption{\label{fig:odf_time_scan} \textbf{Optical dipole force signal and effect of radial cooling.} The ODF signal is the probability to find \Be in the $\ket{\uparrow}$ state after exciting the shared axial out-of-phase motion with an ODF, and probing this excitation with a \Be Raman red sideband $\pi$-pulse. The red data shows the ODF signal with the \nm{1050} beams aligned onto \BeNoSpace, the grey and turquoise data shows the ODF signal for alignment on \HtwopNoSpace. The grey (turquoise) data is without (with) ODF gradient assisted Doppler cooling of the radial modes of motion, demonstrating the reduction of the Debye-Waller effect. Each data point is 300 repetitions of the respective sequence. Error bars indicate the statistical uncertainty (SEM). The solid lines show theoretical predictions for the signal taking into account the corresponding ion's polarizability and Lamb-Dicke parameter which depends on the mode participation of the ion in the axial out-of-phase mode.
}
\end{figure}

\begin{figure}
\includegraphics[width=0.48\textwidth]{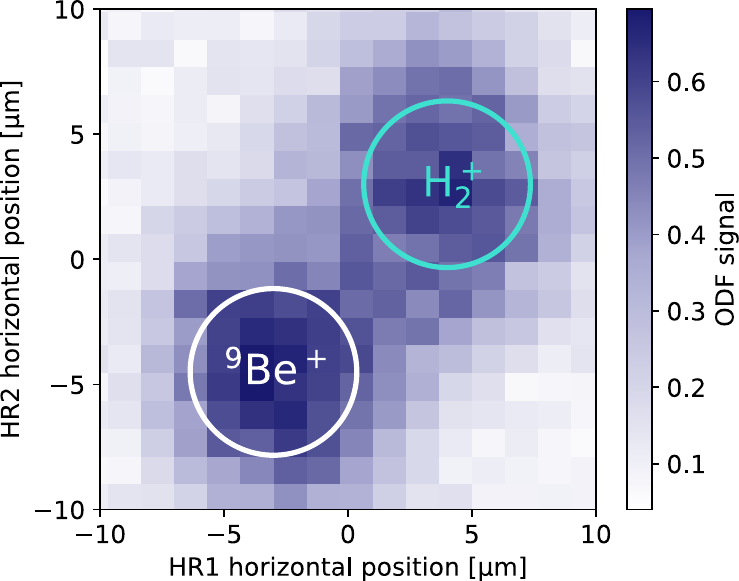}
\caption{\label{fig:odf_two_ions} \textbf{Optical dipole force signal from \boldmath$\mathrm{H}_2^+-{^9\mathrm{Be}}^+$ ion pair.} 
The ODF signal is the probability to find \Be in the $\ket{\uparrow}$ state after exciting the shared motion with an ODF, and probing this excitation with a \Be Raman red sideband $\pi$-pulse. When scanning the position of the HR1 and HR2 beams, we can distinguish the ODF signal from the two individual ions. The data shows a scan of changing the horizontal position of the two laser beams. The circles indicate approximate positions of the two ions to guide the eye.
}
\end{figure}

The \nm{1050} Raman beams are aligned onto the \Htwop ion using an optical dipole force (ODF) experiment \cite{theswagler}. For this we ground-state cool an axial motional mode using the \Be ion. A pulse of the \nm{1050} beams is applied with a difference frequency of the two beams that matches the ground-state cooled motional mode frequency. A red sideband pulse on \Be is driven, followed by fluorescence detection of \BeNoSpace. The \nm{1050} beams (with equal, linear polarization) form a standing wave modulated at the frequency of the motional mode. If the beams are aligned onto either ion the motion is coherently excited and leads to a spin excitation of the \Be ion through the red sideband pulse. 
The ODF strength is proportional to the polarizability $\alpha$, making it easier to find the initial signal on \Be ($\alpha_{\mathrm{Be}} = 26.78 \, \mathrm{a.u.}$ \cite{22UDportal}) than \Htwop ($\alpha_{\mathrm{H2}} = 2.6 - 4.0 \, \mathrm{a.u.}$ for the $\nu=0,N=1$ state), including here the range of the hyper-polarizability variation for the different hyperfine state, which alters the ODF strength). The ODF experiment is initially performed with the axial in-phase of motion and with de-focused beams, roughly aligning one \nm{1050} beam onto the ions by overlapping it with a \nm{313} \Be Raman beam and ensuring overlap of the second, counter-propagating \nm{1050} beam by coupling it into the fiber of the first beam after exiting the vacuum chamber. The axial in-phase mode of motion is used in the initial beam alignment because \Be has a strong mode participation, leading to a stronger ODF. Once this initial signal is found, the focus and position of both beams is optimized iteratively. This leads to the beams being optimally aligned onto the \Be ion. The ODF signal with the beams aligned onto \Be is shown in \hyperref[fig:odf_time_scan]{Figure 6} and agrees well with simulations. To distinguish the signal of \Be and \Htwop we perform the ODF experiment on the axial out-of-phase mode of motion, for which, conveniently, the stronger mode participation of \Htwop compensates its weaker the polarizability, leading to a roughly equal ODF strength for the two ions. With piezo-driven mirrors, we scan the individual beam position to illuminate the individual ions, clearly observing separate signals from \Be and \HtwopNoSpace, see \hyperref[fig:odf_two_ions]{Figure 7}.

Performing the ODF experiment with the beams aligned to the \Htwop ion, we observe a strong decay of the signal, not resembling the theoretical prediction (see \hyperref[fig:odf_time_scan]{Figure 6}). This can be explained by high occupation of the radial modes of motion dominated by \HtwopNoSpace. 
Since the \nm{1050} beams have a projection onto the radial plane (perpendicular to the trap axis), the ODF strength is sensitive to the radial motion through the Debye-Waller effect \cite{98Wineland1}. The two in-phase radial modes of motion of the ion pair only weakly couple to the \Be ion and do not thermalize close to the Doppler limit when Doppler cooling \BeNoSpace. This causes a much stronger Debye-Waller effect for \Htwop than for \BeNoSpace. Peforming the ODF experiment on \Htwop with the radial cooling (see \hyperref[Methods:excooling]{Appendix C}), reduces this effect and results in a similar signal as for \Be (see \hyperref[fig:odf_time_scan]{Figure 6}).

\section*{Appendix C: Optical dipole force gradient assisted Doppler cooling}\label{Methods:excooling} 
The two in-phase radial modes of motion only weakly couple to the \Be ion and do not thermalize close to the Doppler limit when Doppler cooling \BeNoSpace. 
Similar to axialization in Penning traps \cite{1975WinelandAxial}, and more recent demonstrations in Paul traps by Gorman \etal \cite{2014Gorman} and Hou \etal \cite{2024Hou}, we parametrically couple the radial modes to a coolable axial mode. However, due our ion trap size and geometry is not possible to modulate the trap potential to achieve good coupling. Instead, we couple the modes using the gradient of the ODF, tuning the difference frequency of the \nm{1050} beams to the difference of a radial mode and the axial out-of-phase mode. The \nm{1050} beams then implement a population swapping interaction between the addressed motional modes. We alternate the swap operation with Doppler cooling pulses, effectively cooling the radial modes. 
With this additional cooling step the \Htwop ODF achieves similar contrast as for \Be (see \hyperref[fig:odf_time_scan]{Figure 6}). The radial cooling also improves the quality of the Raman-hyperfine transitions in \HtwopNoSpace, which equally suffer from the Debye-Waller effect. King \etal have presented an alternative method to cool the radial motion, using the internal state of the spectroscopy ion to swap population at the single-quantum level between an isolated motional mode and a coolable one \cite{2021King}. This scheme cannot be used here as it would require the \Htwop to be prepared in a hyperfine state when initiating cooling. In our case however, we have to cool the radial motion to enable the state preparation. Our scheme is general and only requires the spectroscopy ion to be sufficiently polarizable by laser light.

\section*{Appendix D: \boldmath$\mathrm{H}_2^+$ hyperfine state preparation}\label{Methods:Pumping}
To prepare the hyperfine state of \Htwop in a pure quantum state, we combine two QLS pumping steps ($\ket{C} \rightarrow \ket{D} \rightarrow \ket{T}$) with many microwave transitions, to transfer population from the entire hyperfine structure to \ket{C}.
Frequency values for the microwave pulses in the state preparation are calculated by numerical diagonalization of the combined hyperfine and Zeeman Hamiltonian at our chosen bias magnetic field of \qty{450}{\micro T} \cite{korobov2006, karr2008, 08Karr} using the recently improved theoretical hyperfine interaction coefficients \cite{2020KarrPRA, haidar2022}.
Microwave $\pi$-times were measured for three microwave transitions with $\Delta m_J = -1,0,1$, respectively, and are calculated for all other transitions by scaling the $\pi$-times with the ratios of the calculated transition dipole moments. Frequencies of the Raman motional sideband transitions are calibrated regularly to compensate for drifts of the axial out-of-phase mode frequency and the laser beam intensity causing AC Stark shifts. 

To improve the fidelity of the state preparation we repeat pulses involved in QLS in a specific manner. We drive a blue-sideband $\pi$-pulse on the $\ket D \rightarrow \ket T$ transition of \HtwopNoSpace, followed by a block of a red-sideband $\pi$-pulse and a spin-state repumping sequence on \BeNoSpace that is repeated twice. This set of pulses is repeated a second time to complete a single QLS pumping step $\ket D \rightarrow \ket T$. The same sequence is used for $\ket C \rightarrow \ket D$. To pump the population from any hyperfine state to $\ket T$, the sequence is as follows: 
State-specific sequence of microwave $\pi$-pulses to transfer the population to \ket{C} , $\ket C \rightarrow \ket D$, $\ket D \rightarrow \ket T$, $\ket C \rightarrow \ket D$, $\ket D \rightarrow \ket T$.

\begin{figure}
\includegraphics[width=0.48\textwidth]{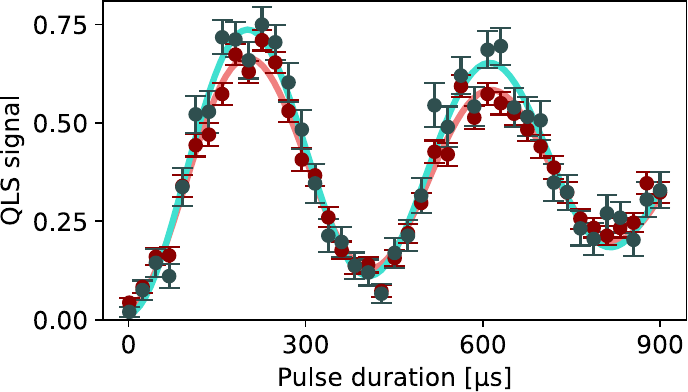}
\caption{\label{fig:heralded} \textbf{Heralded quantum logic spectroscopy signal of ortho-\boldmath$\mathrm{H}_2^+$.} The QLS signal is the probability to find \Be in the $\ket{\uparrow}$ state after the QLS readout and is proportional to the \ket{T} population before QLS readout. The red data is the full data taking all 300 repetitions into account, as presented in \hyperref[fig:alldata]{Figure 4.b}. The turquoise data is post-selected based on the herald detection. This leads to an average of $\approx 100$ repetitions being considered in each data point. Error bars indicate the statistical uncertainty (SEM) for the corresponding repetitions per data point. Solid lines are fits to the data (cosine functions with exponential damping). The contrast is improved from 0.665(8) to 0.74(1) by post selecting the raw data.}
\end{figure}
The required microwave pulse sequences for addressing the entire hyperfine structure are found using a shortest path algorithm on a directed graph, where the nodes represent populations among the hyperfine states and weighted edges represent possible operations and their duration. The hyperfine states are encoded as binary vectors where a 1(0) indicates non-zero(zero) population present. Microwave $\pi$-pulses can swap the population of two states and therefore swap a 0 and a 1. A QLS pumping step can combine two populations into one, combining two 1's into a single 1 and a new 0. Initially, the algorithm assumes population everywhere (all 1's). The algorithm finds the shortest path to the graph node where the target state is the only one populated. More details can be found in \cite{23Stadler}.

The \Htwop hyperfine state preparation sequence consists of $\approx 100$ pulses on \Htwop and $\approx 400$ pulses on \BeNoSpace, with a total duration of $\SI{65}{ms}$. All Doppler cooling, ground state cooling, \Be state preparation and fluorescent readout pulses described above take an additional \SI{85}{ms}, resulting in a full sequence length of \SI{150}{ms}.

\section*{Appendix E: Heralding and post-selection}\label{Methods:Herald} 
In the data presented in the main text, we do not perform any post-selection. We can, however, perform a heralding measurement after the state-preparation, and before the actual measurement, that makes it possible to select only the data points where the \Htwop ion was in the correct hyperfine state. 
The unheralded data is shown in \hyperref[fig:alldata]{Figure 4.b} and again in \hyperref[fig:heralded]{Figure 8} in red. The herald flags if the $\ket T$ state was successfully prepared in each experimental run. We post-select the data based on `positive' herald detections. This results in a post-selected state-preparation fidelity. The post-selection increases the contrast from 0.665(8) to 0.74(1). The heralded data is shown in \hyperref[fig:heralded]{Figure 8} in turquoise.

The heralding measurement drives population from \ket{T} to \ket{D}. Therefore, we reset the hyperfine state to \ket{T} by performing QLS pumping on $\ket{D} \rightarrow \ket T$ after the heralding measurement and before the actual measurement.

\bibliography{refs}

\begin{thebibliography}{57}%
\makeatletter
\providecommand \@ifxundefined [1]{%
 \@ifx{#1\undefined}
}%
\providecommand \@ifnum [1]{%
 \ifnum #1\expandafter \@firstoftwo
 \else \expandafter \@secondoftwo
 \fi
}%
\providecommand \@ifx [1]{%
 \ifx #1\expandafter \@firstoftwo
 \else \expandafter \@secondoftwo
 \fi
}%
\providecommand \natexlab [1]{#1}%
\providecommand \enquote  [1]{``#1''}%
\providecommand \bibnamefont  [1]{#1}%
\providecommand \bibfnamefont [1]{#1}%
\providecommand \citenamefont [1]{#1}%
\providecommand \href@noop [0]{\@secondoftwo}%
\providecommand \href [0]{\begingroup \@sanitize@url \@href}%
\providecommand \@href[1]{\@@startlink{#1}\@@href}%
\providecommand \@@href[1]{\endgroup#1\@@endlink}%
\providecommand \@sanitize@url [0]{\catcode `\\12\catcode `\$12\catcode `\&12\catcode `\#12\catcode `\^12\catcode `\_12\catcode `\%12\relax}%
\providecommand \@@startlink[1]{}%
\providecommand \@@endlink[0]{}%
\providecommand \url  [0]{\begingroup\@sanitize@url \@url }%
\providecommand \@url [1]{\endgroup\@href {#1}{\urlprefix }}%
\providecommand \urlprefix  [0]{URL }%
\providecommand \Eprint [0]{\href }%
\providecommand \doibase [0]{https://doi.org/}%
\providecommand \selectlanguage [0]{\@gobble}%
\providecommand \bibinfo  [0]{\@secondoftwo}%
\providecommand \bibfield  [0]{\@secondoftwo}%
\providecommand \translation [1]{[#1]}%
\providecommand \BibitemOpen [0]{}%
\providecommand \bibitemStop [0]{}%
\providecommand \bibitemNoStop [0]{.\EOS\space}%
\providecommand \EOS [0]{\spacefactor3000\relax}%
\providecommand \BibitemShut  [1]{\csname bibitem#1\endcsname}%
\let\auto@bib@innerbib\@empty
\bibitem [{\citenamefont {Korobov}\ \emph {et~al.}(2017)\citenamefont {Korobov}, \citenamefont {Hilico},\ and\ \citenamefont {Karr}}]{17Korobov}%
  \BibitemOpen
  \bibfield  {author} {\bibinfo {author} {\bibfnamefont {V.~I.}\ \bibnamefont {Korobov}}, \bibinfo {author} {\bibfnamefont {L.}~\bibnamefont {Hilico}},\ and\ \bibinfo {author} {\bibfnamefont {J.-P.}\ \bibnamefont {Karr}},\ }\bibfield  {title} {\bibinfo {title} {Fundamental transitions and ionization energies of the hydrogen molecular ions with few ppt uncertainty},\ }\href {https://doi.org/10.1103/PhysRevLett.118.233001} {\bibfield  {journal} {\bibinfo  {journal} {Phys. Rev. Lett.}\ }\textbf {\bibinfo {volume} {118}},\ \bibinfo {pages} {233001} (\bibinfo {year} {2017})}\BibitemShut {NoStop}%
\bibitem [{\citenamefont {Schiller}(2022)}]{2022Schiller}%
  \BibitemOpen
  \bibfield  {author} {\bibinfo {author} {\bibfnamefont {S.}~\bibnamefont {Schiller}},\ }\bibfield  {title} {\bibinfo {title} {Precision spectroscopy of molecular hydrogen ions: an introduction},\ }\href {https://doi.org/10.1080/00107514.2023.2180180} {\bibfield  {journal} {\bibinfo  {journal} {Contemporary Physics}\ }\textbf {\bibinfo {volume} {63}},\ \bibinfo {pages} {247} (\bibinfo {year} {2022})}\BibitemShut {NoStop}%
\bibitem [{\citenamefont {Delaunay}\ \emph {et~al.}(2023)\citenamefont {Delaunay}, \citenamefont {Karr}, \citenamefont {Kitahara}, \citenamefont {Koelemeij}, \citenamefont {Soreq},\ and\ \citenamefont {Zupan}}]{2023Delaunay}%
  \BibitemOpen
  \bibfield  {author} {\bibinfo {author} {\bibfnamefont {C.}~\bibnamefont {Delaunay}}, \bibinfo {author} {\bibfnamefont {J.-P.}\ \bibnamefont {Karr}}, \bibinfo {author} {\bibfnamefont {T.}~\bibnamefont {Kitahara}}, \bibinfo {author} {\bibfnamefont {J.~C.~J.}\ \bibnamefont {Koelemeij}}, \bibinfo {author} {\bibfnamefont {Y.}~\bibnamefont {Soreq}},\ and\ \bibinfo {author} {\bibfnamefont {J.}~\bibnamefont {Zupan}},\ }\bibfield  {title} {\bibinfo {title} {Self-consistent extraction of spectroscopic bounds on light new physics},\ }\href {https://doi.org/10.1103/PhysRevLett.130.121801} {\bibfield  {journal} {\bibinfo  {journal} {Phys. Rev. Lett.}\ }\textbf {\bibinfo {volume} {130}},\ \bibinfo {pages} {121801} (\bibinfo {year} {2023})}\BibitemShut {NoStop}%
\bibitem [{\citenamefont {Schiller}\ and\ \citenamefont {Karr}(2024{\natexlab{a}})}]{2024schillerKarr}%
  \BibitemOpen
  \bibfield  {author} {\bibinfo {author} {\bibfnamefont {S.}~\bibnamefont {Schiller}}\ and\ \bibinfo {author} {\bibfnamefont {J.-P.}\ \bibnamefont {Karr}},\ }\bibfield  {title} {\bibinfo {title} {Prospects for the determination of fundamental constants with beyond-state-of-the-art uncertainty using molecular hydrogen ion spectroscopy},\ }\href {https://doi.org/10.1103/PhysRevA.109.042825} {\bibfield  {journal} {\bibinfo  {journal} {Phys. Rev. A}\ }\textbf {\bibinfo {volume} {109}},\ \bibinfo {pages} {042825} (\bibinfo {year} {2024}{\natexlab{a}})}\BibitemShut {NoStop}%
\bibitem [{\citenamefont {Schneider}\ \emph {et~al.}(2010)\citenamefont {Schneider}, \citenamefont {Roth}, \citenamefont {Duncker}, \citenamefont {Ernsting},\ and\ \citenamefont {Schiller}}]{10Schneider2}%
  \BibitemOpen
  \bibfield  {author} {\bibinfo {author} {\bibfnamefont {T.}~\bibnamefont {Schneider}}, \bibinfo {author} {\bibfnamefont {B.}~\bibnamefont {Roth}}, \bibinfo {author} {\bibfnamefont {H.}~\bibnamefont {Duncker}}, \bibinfo {author} {\bibfnamefont {I.}~\bibnamefont {Ernsting}},\ and\ \bibinfo {author} {\bibfnamefont {S.}~\bibnamefont {Schiller}},\ }\bibfield  {title} {\bibinfo {title} {All-optical preparation of molecular ions in the rovibrational ground state},\ }\href {https://doi.org/10.1038/nphys1605} {\bibfield  {journal} {\bibinfo  {journal} {Nature Physics}\ }\textbf {\bibinfo {volume} {6}},\ \bibinfo {pages} {275} (\bibinfo {year} {2010})}\BibitemShut {NoStop}%
\bibitem [{\citenamefont {Doran}\ \emph {et~al.}(2024)\citenamefont {Doran}, \citenamefont {H\"olsch}, \citenamefont {Beyer},\ and\ \citenamefont {Merkt}}]{2024Doran}%
  \BibitemOpen
  \bibfield  {author} {\bibinfo {author} {\bibfnamefont {I.}~\bibnamefont {Doran}}, \bibinfo {author} {\bibfnamefont {N.}~\bibnamefont {H\"olsch}}, \bibinfo {author} {\bibfnamefont {M.}~\bibnamefont {Beyer}},\ and\ \bibinfo {author} {\bibfnamefont {F.}~\bibnamefont {Merkt}},\ }\bibfield  {title} {\bibinfo {title} {Zero-quantum-defect method and the fundamental vibrational interval of \ensuremath{\mathrm{\uppercase{h}}_2^+}},\ }\href {https://doi.org/10.1103/PhysRevLett.132.073001} {\bibfield  {journal} {\bibinfo  {journal} {Phys. Rev. Lett.}\ }\textbf {\bibinfo {volume} {132}},\ \bibinfo {pages} {073001} (\bibinfo {year} {2024})}\BibitemShut {NoStop}%
\bibitem [{\citenamefont {Schenkel}\ \emph {et~al.}(2024)\citenamefont {Schenkel}, \citenamefont {Alighanbari},\ and\ \citenamefont {Schiller}}]{2024Schenkel}%
  \BibitemOpen
  \bibfield  {author} {\bibinfo {author} {\bibfnamefont {M.~R.}\ \bibnamefont {Schenkel}}, \bibinfo {author} {\bibfnamefont {S.}~\bibnamefont {Alighanbari}},\ and\ \bibinfo {author} {\bibfnamefont {S.}~\bibnamefont {Schiller}},\ }\bibfield  {title} {\bibinfo {title} {Laser spectroscopy of a rovibrational transition in the molecular hydrogen ion \ensuremath{\mathrm{\uppercase{h}}_2^+}},\ }\href {https://doi.org/10.1038/s41567-023-02320-z} {\bibfield  {journal} {\bibinfo  {journal} {Nature Physics}\ }\textbf {\bibinfo {volume} {20}},\ \bibinfo {pages} {383} (\bibinfo {year} {2024})}\BibitemShut {NoStop}%
\bibitem [{\citenamefont {Huntemann}\ \emph {et~al.}(2016)\citenamefont {Huntemann}, \citenamefont {Sanner}, \citenamefont {Lipphardt}, \citenamefont {Tamm},\ and\ \citenamefont {Peik}}]{16Huntemann}%
  \BibitemOpen
  \bibfield  {author} {\bibinfo {author} {\bibfnamefont {N.}~\bibnamefont {Huntemann}}, \bibinfo {author} {\bibfnamefont {C.}~\bibnamefont {Sanner}}, \bibinfo {author} {\bibfnamefont {B.}~\bibnamefont {Lipphardt}}, \bibinfo {author} {\bibfnamefont {C.}~\bibnamefont {Tamm}},\ and\ \bibinfo {author} {\bibfnamefont {E.}~\bibnamefont {Peik}},\ }\bibfield  {title} {\bibinfo {title} {Single-ion atomic clock with $3\ifmmode\times\else\texttimes\fi{}{10}^{\ensuremath{-}18}$ systematic uncertainty},\ }\href {https://doi.org/10.1103/PhysRevLett.116.063001} {\bibfield  {journal} {\bibinfo  {journal} {Phys. Rev. Lett.}\ }\textbf {\bibinfo {volume} {116}},\ \bibinfo {pages} {063001} (\bibinfo {year} {2016})}\BibitemShut {NoStop}%
\bibitem [{\citenamefont {Brewer}\ \emph {et~al.}(2019)\citenamefont {Brewer}, \citenamefont {Chen}, \citenamefont {Hankin}, \citenamefont {Clements}, \citenamefont {Chou}, \citenamefont {Wineland}, \citenamefont {Hume},\ and\ \citenamefont {Leibrandt}}]{19Brewer}%
  \BibitemOpen
  \bibfield  {author} {\bibinfo {author} {\bibfnamefont {S.~M.}\ \bibnamefont {Brewer}}, \bibinfo {author} {\bibfnamefont {J.-S.}\ \bibnamefont {Chen}}, \bibinfo {author} {\bibfnamefont {A.~M.}\ \bibnamefont {Hankin}}, \bibinfo {author} {\bibfnamefont {E.~R.}\ \bibnamefont {Clements}}, \bibinfo {author} {\bibfnamefont {C.-W.}\ \bibnamefont {Chou}}, \bibinfo {author} {\bibfnamefont {D.~J.}\ \bibnamefont {Wineland}}, \bibinfo {author} {\bibfnamefont {D.~B.}\ \bibnamefont {Hume}},\ and\ \bibinfo {author} {\bibfnamefont {D.~R.}\ \bibnamefont {Leibrandt}},\ }\bibfield  {title} {\bibinfo {title} {{$^{27}{\mathrm{Al}}^{+}$} quantum-logic clock with a systematic uncertainty below ${10}^{\ensuremath{-}18}$},\ }\href {https://doi.org/10.1103/PhysRevLett.123.033201} {\bibfield  {journal} {\bibinfo  {journal} {Phys. Rev. Lett.}\ }\textbf {\bibinfo {volume} {123}},\ \bibinfo {pages} {033201} (\bibinfo {year} {2019})}\BibitemShut {NoStop}%
\bibitem [{\citenamefont {Bakalov}\ and\ \citenamefont {Schiller}(2014)}]{2014Bakalov}%
  \BibitemOpen
  \bibfield  {author} {\bibinfo {author} {\bibfnamefont {D.}~\bibnamefont {Bakalov}}\ and\ \bibinfo {author} {\bibfnamefont {S.}~\bibnamefont {Schiller}},\ }\bibfield  {title} {\bibinfo {title} {The electric quadrupole moment of molecular hydrogen ions and their potential for a molecular ion clock},\ }\href {https://doi.org/10.1007/s00340-013-5703-z} {\bibfield  {journal} {\bibinfo  {journal} {Applied Physics B}\ }\textbf {\bibinfo {volume} {114}},\ \bibinfo {pages} {213} (\bibinfo {year} {2014})}\BibitemShut {NoStop}%
\bibitem [{\citenamefont {Schiller}\ \emph {et~al.}(2014)\citenamefont {Schiller}, \citenamefont {Bakalov},\ and\ \citenamefont {Korobov}}]{14Schiller}%
  \BibitemOpen
  \bibfield  {author} {\bibinfo {author} {\bibfnamefont {S.}~\bibnamefont {Schiller}}, \bibinfo {author} {\bibfnamefont {D.}~\bibnamefont {Bakalov}},\ and\ \bibinfo {author} {\bibfnamefont {V.~I.}\ \bibnamefont {Korobov}},\ }\bibfield  {title} {\bibinfo {title} {Simplest molecules as candidates for precise optical clocks},\ }\href {https://doi.org/10.1103/PhysRevLett.113.023004} {\bibfield  {journal} {\bibinfo  {journal} {Phys. Rev. Lett.}\ }\textbf {\bibinfo {volume} {113}},\ \bibinfo {pages} {023004} (\bibinfo {year} {2014})}\BibitemShut {NoStop}%
\bibitem [{\citenamefont {Karr}\ \emph {et~al.}(2016)\citenamefont {Karr}, \citenamefont {Patra}, \citenamefont {Koelemeij}, \citenamefont {Heinrich}, \citenamefont {Sillitoe}, \citenamefont {Douillet},\ and\ \citenamefont {Hilico}}]{16Karr1}%
  \BibitemOpen
  \bibfield  {author} {\bibinfo {author} {\bibfnamefont {J.-P.}\ \bibnamefont {Karr}}, \bibinfo {author} {\bibfnamefont {S.}~\bibnamefont {Patra}}, \bibinfo {author} {\bibfnamefont {J.~C.~J.}\ \bibnamefont {Koelemeij}}, \bibinfo {author} {\bibfnamefont {J.}~\bibnamefont {Heinrich}}, \bibinfo {author} {\bibfnamefont {N.}~\bibnamefont {Sillitoe}}, \bibinfo {author} {\bibfnamefont {A.}~\bibnamefont {Douillet}},\ and\ \bibinfo {author} {\bibfnamefont {L.}~\bibnamefont {Hilico}},\ }\bibfield  {title} {\bibinfo {title} {Hydrogen molecular ions: new schemes for metrology and fundamental physics tests},\ }\href {http://stacks.iop.org/1742-6596/723/i=1/a=012048} {\bibfield  {journal} {\bibinfo  {journal} {Journal of Physics: Conference Series}\ }\textbf {\bibinfo {volume} {723}},\ \bibinfo {pages} {012048} (\bibinfo {year} {2016})}\BibitemShut {NoStop}%
\bibitem [{\citenamefont {Wineland}\ \emph {et~al.}(1998)\citenamefont {Wineland}, \citenamefont {Monroe}, \citenamefont {Itano}, \citenamefont {Leibfried}, \citenamefont {King},\ and\ \citenamefont {Meekhof}}]{98Wineland1}%
  \BibitemOpen
  \bibfield  {author} {\bibinfo {author} {\bibfnamefont {D.~J.}\ \bibnamefont {Wineland}}, \bibinfo {author} {\bibfnamefont {C.}~\bibnamefont {Monroe}}, \bibinfo {author} {\bibfnamefont {W.~M.}\ \bibnamefont {Itano}}, \bibinfo {author} {\bibfnamefont {D.}~\bibnamefont {Leibfried}}, \bibinfo {author} {\bibfnamefont {B.~E.}\ \bibnamefont {King}},\ and\ \bibinfo {author} {\bibfnamefont {D.~M.}\ \bibnamefont {Meekhof}},\ }\bibfield  {title} {\bibinfo {title} {Experimental issues in coherent quantum-state manipulation of trapped atomic ions},\ }\href {https://www.ncbi.nlm.nih.gov/pmc/articles/PMC4898965/} {\bibfield  {journal} {\bibinfo  {journal} {J. Res. Natl. Inst. Stand. Technol.}\ }\textbf {\bibinfo {volume} {103}},\ \bibinfo {pages} {259} (\bibinfo {year} {1998})}\BibitemShut {NoStop}%
\bibitem [{\citenamefont {Schmidt}\ \emph {et~al.}(2005)\citenamefont {Schmidt}, \citenamefont {Rosenband}, \citenamefont {Langer}, \citenamefont {Itano}, \citenamefont {Bergquist},\ and\ \citenamefont {Wineland}}]{05Schmidt}%
  \BibitemOpen
  \bibfield  {author} {\bibinfo {author} {\bibfnamefont {P.~O.}\ \bibnamefont {Schmidt}}, \bibinfo {author} {\bibfnamefont {T.}~\bibnamefont {Rosenband}}, \bibinfo {author} {\bibfnamefont {C.}~\bibnamefont {Langer}}, \bibinfo {author} {\bibfnamefont {W.~M.}\ \bibnamefont {Itano}}, \bibinfo {author} {\bibfnamefont {J.~C.}\ \bibnamefont {Bergquist}},\ and\ \bibinfo {author} {\bibfnamefont {D.~J.}\ \bibnamefont {Wineland}},\ }\bibfield  {title} {\bibinfo {title} {Spectroscopy using quantum logic},\ }\href {https://doi.org/10.1126/science.1114375} {\bibfield  {journal} {\bibinfo  {journal} {Science}\ }\textbf {\bibinfo {volume} {309}},\ \bibinfo {pages} {749} (\bibinfo {year} {2005})}\BibitemShut {NoStop}%
\bibitem [{\citenamefont {Wolf}\ \emph {et~al.}(2016)\citenamefont {Wolf}, \citenamefont {Wan}, \citenamefont {Heip}, \citenamefont {Gebert}, \citenamefont {Shi},\ and\ \citenamefont {Schmidt}}]{16Wolf}%
  \BibitemOpen
  \bibfield  {author} {\bibinfo {author} {\bibfnamefont {F.}~\bibnamefont {Wolf}}, \bibinfo {author} {\bibfnamefont {Y.}~\bibnamefont {Wan}}, \bibinfo {author} {\bibfnamefont {J.~C.}\ \bibnamefont {Heip}}, \bibinfo {author} {\bibfnamefont {F.}~\bibnamefont {Gebert}}, \bibinfo {author} {\bibfnamefont {C.}~\bibnamefont {Shi}},\ and\ \bibinfo {author} {\bibfnamefont {P.~O.}\ \bibnamefont {Schmidt}},\ }\bibfield  {title} {\bibinfo {title} {{Non-destructive state detection for quantum logic spectroscopy of molecular ions}},\ }\href {https://www.nature.com/articles/nature16513} {\bibfield  {journal} {\bibinfo  {journal} {Nature}\ }\textbf {\bibinfo {volume} {530}},\ \bibinfo {pages} {457} (\bibinfo {year} {2016})}\BibitemShut {NoStop}%
\bibitem [{\citenamefont {Chou}\ \emph {et~al.}(2017)\citenamefont {Chou}, \citenamefont {Kurz}, \citenamefont {Hume}, \citenamefont {Plessow}, \citenamefont {Leibrandt},\ and\ \citenamefont {Leibfried}}]{17Chou}%
  \BibitemOpen
  \bibfield  {author} {\bibinfo {author} {\bibfnamefont {C.-W.}\ \bibnamefont {Chou}}, \bibinfo {author} {\bibfnamefont {C.}~\bibnamefont {Kurz}}, \bibinfo {author} {\bibfnamefont {D.~B.}\ \bibnamefont {Hume}}, \bibinfo {author} {\bibfnamefont {P.~N.}\ \bibnamefont {Plessow}}, \bibinfo {author} {\bibfnamefont {D.~R.}\ \bibnamefont {Leibrandt}},\ and\ \bibinfo {author} {\bibfnamefont {D.}~\bibnamefont {Leibfried}},\ }\bibfield  {title} {\bibinfo {title} {Preparation and coherent manipulation of pure quantum states of a single molecular ion},\ }\href {http://dx.doi.org/10.1038/nature22338} {\bibfield  {journal} {\bibinfo  {journal} {Nature}\ }\textbf {\bibinfo {volume} {545}},\ \bibinfo {pages} {203} (\bibinfo {year} {2017})}\BibitemShut {NoStop}%
\bibitem [{\citenamefont {Sinhal}\ \emph {et~al.}(2020)\citenamefont {Sinhal}, \citenamefont {Meir}, \citenamefont {Najafian}, \citenamefont {Hegi},\ and\ \citenamefont {Willitsch}}]{2020Sinhal}%
  \BibitemOpen
  \bibfield  {author} {\bibinfo {author} {\bibfnamefont {M.}~\bibnamefont {Sinhal}}, \bibinfo {author} {\bibfnamefont {Z.}~\bibnamefont {Meir}}, \bibinfo {author} {\bibfnamefont {K.}~\bibnamefont {Najafian}}, \bibinfo {author} {\bibfnamefont {G.}~\bibnamefont {Hegi}},\ and\ \bibinfo {author} {\bibfnamefont {S.}~\bibnamefont {Willitsch}},\ }\bibfield  {title} {\bibinfo {title} {Quantum-nondemolition state detection and spectroscopy of single trapped molecules},\ }\href {https://doi.org/10.1126/science.aaz9837} {\bibfield  {journal} {\bibinfo  {journal} {Science}\ }\textbf {\bibinfo {volume} {367}},\ \bibinfo {pages} {1213} (\bibinfo {year} {2020})}\BibitemShut {NoStop}%
\bibitem [{\citenamefont {Chou}\ \emph {et~al.}(2020)\citenamefont {Chou}, \citenamefont {Collopy}, \citenamefont {Kurz}, \citenamefont {Lin}, \citenamefont {Harding}, \citenamefont {Plessow}, \citenamefont {Fortier}, \citenamefont {Diddams}, \citenamefont {Leibfried},\ and\ \citenamefont {Leibrandt}}]{2020Chou}%
  \BibitemOpen
  \bibfield  {author} {\bibinfo {author} {\bibfnamefont {C.-W.}\ \bibnamefont {Chou}}, \bibinfo {author} {\bibfnamefont {A.~L.}\ \bibnamefont {Collopy}}, \bibinfo {author} {\bibfnamefont {C.}~\bibnamefont {Kurz}}, \bibinfo {author} {\bibfnamefont {Y.}~\bibnamefont {Lin}}, \bibinfo {author} {\bibfnamefont {M.~E.}\ \bibnamefont {Harding}}, \bibinfo {author} {\bibfnamefont {P.~N.}\ \bibnamefont {Plessow}}, \bibinfo {author} {\bibfnamefont {T.}~\bibnamefont {Fortier}}, \bibinfo {author} {\bibfnamefont {S.}~\bibnamefont {Diddams}}, \bibinfo {author} {\bibfnamefont {D.}~\bibnamefont {Leibfried}},\ and\ \bibinfo {author} {\bibfnamefont {D.~R.}\ \bibnamefont {Leibrandt}},\ }\bibfield  {title} {\bibinfo {title} {Frequency-comb spectroscopy on pure quantum states of a single molecular ion},\ }\href {https://doi.org/10.1126/science.aba3628} {\bibfield  {journal} {\bibinfo  {journal} {Science}\ }\textbf {\bibinfo {volume} {367}},\ \bibinfo {pages} {1458} (\bibinfo {year} {2020})}\BibitemShut {NoStop}%
\bibitem [{\citenamefont {Lin}\ \emph {et~al.}(2020)\citenamefont {Lin}, \citenamefont {Leibrandt}, \citenamefont {Leibfried},\ and\ \citenamefont {Chou}}]{2020Lin}%
  \BibitemOpen
  \bibfield  {author} {\bibinfo {author} {\bibfnamefont {Y.}~\bibnamefont {Lin}}, \bibinfo {author} {\bibfnamefont {D.~R.}\ \bibnamefont {Leibrandt}}, \bibinfo {author} {\bibfnamefont {D.}~\bibnamefont {Leibfried}},\ and\ \bibinfo {author} {\bibfnamefont {C.-W.}\ \bibnamefont {Chou}},\ }\bibfield  {title} {\bibinfo {title} {Quantum entanglement between an atom and a molecule},\ }\href {https://doi.org/10.1038/s41586-020-2257-1} {\bibfield  {journal} {\bibinfo  {journal} {Nature}\ }\textbf {\bibinfo {volume} {581}},\ \bibinfo {pages} {273} (\bibinfo {year} {2020})}\BibitemShut {NoStop}%
\bibitem [{\citenamefont {Collopy}\ \emph {et~al.}(2023)\citenamefont {Collopy}, \citenamefont {Schmidt}, \citenamefont {Leibfried}, \citenamefont {Leibrandt},\ and\ \citenamefont {Chou}}]{2023Collopy}%
  \BibitemOpen
  \bibfield  {author} {\bibinfo {author} {\bibfnamefont {A.~L.}\ \bibnamefont {Collopy}}, \bibinfo {author} {\bibfnamefont {J.}~\bibnamefont {Schmidt}}, \bibinfo {author} {\bibfnamefont {D.}~\bibnamefont {Leibfried}}, \bibinfo {author} {\bibfnamefont {D.~R.}\ \bibnamefont {Leibrandt}},\ and\ \bibinfo {author} {\bibfnamefont {C.-W.}\ \bibnamefont {Chou}},\ }\bibfield  {title} {\bibinfo {title} {Effects of an oscillating electric field on and dipole moment measurement of a single molecular ion},\ }\href {https://doi.org/10.1103/PhysRevLett.130.223201} {\bibfield  {journal} {\bibinfo  {journal} {Phys. Rev. Lett.}\ }\textbf {\bibinfo {volume} {130}},\ \bibinfo {pages} {223201} (\bibinfo {year} {2023})}\BibitemShut {NoStop}%
\bibitem [{\citenamefont {Liu}\ \emph {et~al.}(2024)\citenamefont {Liu}, \citenamefont {Schmidt}, \citenamefont {Liu}, \citenamefont {Leibrandt}, \citenamefont {Leibfried},\ and\ \citenamefont {wen Chou}}]{2024Liu}%
  \BibitemOpen
  \bibfield  {author} {\bibinfo {author} {\bibfnamefont {Y.}~\bibnamefont {Liu}}, \bibinfo {author} {\bibfnamefont {J.}~\bibnamefont {Schmidt}}, \bibinfo {author} {\bibfnamefont {Z.}~\bibnamefont {Liu}}, \bibinfo {author} {\bibfnamefont {D.~R.}\ \bibnamefont {Leibrandt}}, \bibinfo {author} {\bibfnamefont {D.}~\bibnamefont {Leibfried}},\ and\ \bibinfo {author} {\bibfnamefont {C.}~\bibnamefont {wen Chou}},\ }\bibfield  {title} {\bibinfo {title} {Quantum state tracking and control of a single molecular ion in a thermal environment},\ }\href {https://doi.org/10.1126/science.ado1001} {\bibfield  {journal} {\bibinfo  {journal} {Science}\ }\textbf {\bibinfo {volume} {385}},\ \bibinfo {pages} {790} (\bibinfo {year} {2024})}\BibitemShut {NoStop}%
\bibitem [{\citenamefont {Schwegler}\ \emph {et~al.}(2023)\citenamefont {Schwegler}, \citenamefont {Holzapfel}, \citenamefont {Stadler}, \citenamefont {Mitjans}, \citenamefont {Sergachev}, \citenamefont {Home},\ and\ \citenamefont {Kienzler}}]{2023Schwegler}%
  \BibitemOpen
  \bibfield  {author} {\bibinfo {author} {\bibfnamefont {N.}~\bibnamefont {Schwegler}}, \bibinfo {author} {\bibfnamefont {D.}~\bibnamefont {Holzapfel}}, \bibinfo {author} {\bibfnamefont {M.}~\bibnamefont {Stadler}}, \bibinfo {author} {\bibfnamefont {A.}~\bibnamefont {Mitjans}}, \bibinfo {author} {\bibfnamefont {I.}~\bibnamefont {Sergachev}}, \bibinfo {author} {\bibfnamefont {J.~P.}\ \bibnamefont {Home}},\ and\ \bibinfo {author} {\bibfnamefont {D.}~\bibnamefont {Kienzler}},\ }\bibfield  {title} {\bibinfo {title} {Trapping and ground-state cooling of a single \ensuremath{\mathrm{\uppercase{h}}_2^+}},\ }\href {https://doi.org/10.1103/PhysRevLett.131.133003} {\bibfield  {journal} {\bibinfo  {journal} {Phys. Rev. Lett.}\ }\textbf {\bibinfo {volume} {131}},\ \bibinfo {pages} {133003} (\bibinfo {year} {2023})}\BibitemShut {NoStop}%
\bibitem [{\citenamefont {Weijun}\ \emph {et~al.}(1993)\citenamefont {Weijun}, \citenamefont {Alheit},\ and\ \citenamefont {Werth}}]{93Weijun}%
  \BibitemOpen
  \bibfield  {author} {\bibinfo {author} {\bibfnamefont {Y.}~\bibnamefont {Weijun}}, \bibinfo {author} {\bibfnamefont {R.}~\bibnamefont {Alheit}},\ and\ \bibinfo {author} {\bibfnamefont {G.}~\bibnamefont {Werth}},\ }\bibfield  {title} {\bibinfo {title} {Vibrational population of $\mathrm{H}_2^+$ after electroionization of thermal $\mathrm{H}_2$},\ }\href {https://doi.org/10.1007/BF01436971} {\bibfield  {journal} {\bibinfo  {journal} {Zeitschrift f{\"u}r Physik D Atoms, Molecules and Clusters}\ }\textbf {\bibinfo {volume} {28}},\ \bibinfo {pages} {87} (\bibinfo {year} {1993})}\BibitemShut {NoStop}%
\bibitem [{\citenamefont {Posen}\ \emph {et~al.}(1983)\citenamefont {Posen}, \citenamefont {Dalgarno},\ and\ \citenamefont {Peek}}]{1983Posen}%
  \BibitemOpen
  \bibfield  {author} {\bibinfo {author} {\bibfnamefont {A.}~\bibnamefont {Posen}}, \bibinfo {author} {\bibfnamefont {A.}~\bibnamefont {Dalgarno}},\ and\ \bibinfo {author} {\bibfnamefont {J.}~\bibnamefont {Peek}},\ }\bibfield  {title} {\bibinfo {title} {The quadrupole vibration-rotation transition probabilities of the molecular hydrogen ion},\ }\href {https://doi.org/https://doi.org/10.1016/0092-640X(83)90017-7} {\bibfield  {journal} {\bibinfo  {journal} {Atomic Data and Nuclear Data Tables}\ }\textbf {\bibinfo {volume} {28}},\ \bibinfo {pages} {265} (\bibinfo {year} {1983})}\BibitemShut {NoStop}%
\bibitem [{\citenamefont {Pil{\'{o}}n}(2013)}]{13Pilon}%
  \BibitemOpen
  \bibfield  {author} {\bibinfo {author} {\bibfnamefont {H.~O.}\ \bibnamefont {Pil{\'{o}}n}},\ }\bibfield  {title} {\bibinfo {title} {Quadrupole transitions in the bound rotational{\textendash}vibrational spectrum of the deuterium molecular ion},\ }\href {https://doi.org/10.1088/0953-4075/46/24/245101} {\bibfield  {journal} {\bibinfo  {journal} {Journal of Physics B: Atomic, Molecular and Optical Physics}\ }\textbf {\bibinfo {volume} {46}},\ \bibinfo {pages} {245101} (\bibinfo {year} {2013})}\BibitemShut {NoStop}%
\bibitem [{\citenamefont {Bishop}\ \emph {et~al.}(1975)\citenamefont {Bishop}, \citenamefont {Shih}, \citenamefont {Beckel}, \citenamefont {Wu},\ and\ \citenamefont {Peek}}]{75Bishop}%
  \BibitemOpen
  \bibfield  {author} {\bibinfo {author} {\bibfnamefont {D.~M.}\ \bibnamefont {Bishop}}, \bibinfo {author} {\bibfnamefont {S.}~\bibnamefont {Shih}}, \bibinfo {author} {\bibfnamefont {C.~L.}\ \bibnamefont {Beckel}}, \bibinfo {author} {\bibfnamefont {F.}~\bibnamefont {Wu}},\ and\ \bibinfo {author} {\bibfnamefont {J.~M.}\ \bibnamefont {Peek}},\ }\bibfield  {title} {\bibinfo {title} {Theoretical study of {H$^{+}_{2}$} spectroscopic properties. iv. adiabatic effects for the $2p\pi_{u}$ and $3d\sigma_{g}$ electronic states},\ }\href {https://doi.org/10.1063/1.431226} {\bibfield  {journal} {\bibinfo  {journal} {The Journal of Chemical Physics}\ }\textbf {\bibinfo {volume} {63}},\ \bibinfo {pages} {4836} (\bibinfo {year} {1975})}\BibitemShut {NoStop}%
\bibitem [{\citenamefont {Korobov}\ and\ \citenamefont {Bakalov}(2023)}]{2023Korobov}%
  \BibitemOpen
  \bibfield  {author} {\bibinfo {author} {\bibfnamefont {V.~I.}\ \bibnamefont {Korobov}}\ and\ \bibinfo {author} {\bibfnamefont {D.}~\bibnamefont {Bakalov}},\ }\bibfield  {title} {\bibinfo {title} {Forbidden ortho-para electric dipole transitions in the {H$_2^+$} ion},\ }\href {https://doi.org/10.1103/PhysRevA.107.022812} {\bibfield  {journal} {\bibinfo  {journal} {Phys. Rev. A}\ }\textbf {\bibinfo {volume} {107}},\ \bibinfo {pages} {022812} (\bibinfo {year} {2023})}\BibitemShut {NoStop}%
\bibitem [{\citenamefont {Hansen}\ \emph {et~al.}(2014)\citenamefont {Hansen}, \citenamefont {Versolato}, \citenamefont {Klosowski}, \citenamefont {Kristensen}, \citenamefont {Gingell}, \citenamefont {Schwarz}, \citenamefont {Windberger}, \citenamefont {Ullrich}, \citenamefont {López-Urrutia},\ and\ \citenamefont {Drewsen}}]{14Hansen}%
  \BibitemOpen
  \bibfield  {author} {\bibinfo {author} {\bibfnamefont {A.~K.}\ \bibnamefont {Hansen}}, \bibinfo {author} {\bibfnamefont {O.~O.}\ \bibnamefont {Versolato}}, \bibinfo {author} {\bibfnamefont {L.}~\bibnamefont {Klosowski}}, \bibinfo {author} {\bibfnamefont {S.~B.}\ \bibnamefont {Kristensen}}, \bibinfo {author} {\bibfnamefont {A.}~\bibnamefont {Gingell}}, \bibinfo {author} {\bibfnamefont {M.}~\bibnamefont {Schwarz}}, \bibinfo {author} {\bibfnamefont {A.}~\bibnamefont {Windberger}}, \bibinfo {author} {\bibfnamefont {J.}~\bibnamefont {Ullrich}}, \bibinfo {author} {\bibfnamefont {J.~R.~C.}\ \bibnamefont {López-Urrutia}},\ and\ \bibinfo {author} {\bibfnamefont {M.}~\bibnamefont {Drewsen}},\ }\bibfield  {title} {\bibinfo {title} {Efficient rotational cooling of coulomb-crystallized molecular ions by a helium buffer gas},\ }\href {https://doi.org/10.1038/nature12996} {\bibfield  {journal} {\bibinfo  {journal} {Nature}\ }\textbf {\bibinfo {volume} {508}},\ \bibinfo {pages} {76} (\bibinfo {year} {2014})}\BibitemShut
  {NoStop}%
\bibitem [{\citenamefont {Schiller}\ \emph {et~al.}(2017)\citenamefont {Schiller}, \citenamefont {Kortunov}, \citenamefont {Hern\'andez~Vera}, \citenamefont {Gianturco},\ and\ \citenamefont {da~Silva}}]{17Schiller}%
  \BibitemOpen
  \bibfield  {author} {\bibinfo {author} {\bibfnamefont {S.}~\bibnamefont {Schiller}}, \bibinfo {author} {\bibfnamefont {I.}~\bibnamefont {Kortunov}}, \bibinfo {author} {\bibfnamefont {M.}~\bibnamefont {Hern\'andez~Vera}}, \bibinfo {author} {\bibfnamefont {F.}~\bibnamefont {Gianturco}},\ and\ \bibinfo {author} {\bibfnamefont {H.}~\bibnamefont {da~Silva}},\ }\bibfield  {title} {\bibinfo {title} {Quantum state preparation of homonuclear molecular ions enabled via a cold buffer gas: An ab initio study for the \ensuremath{\mathrm{\uppercase{h}}_2^+} and the \ensuremath{\mathrm{\uppercase{d}}_2^+} case},\ }\href {https://doi.org/10.1103/PhysRevA.95.043411} {\bibfield  {journal} {\bibinfo  {journal} {Phys. Rev. A}\ }\textbf {\bibinfo {volume} {95}},\ \bibinfo {pages} {043411} (\bibinfo {year} {2017})}\BibitemShut {NoStop}%
\bibitem [{\citenamefont {Hankin}\ \emph {et~al.}(2019)\citenamefont {Hankin}, \citenamefont {Clements}, \citenamefont {Huang}, \citenamefont {Brewer}, \citenamefont {Chen}, \citenamefont {Chou}, \citenamefont {Hume},\ and\ \citenamefont {Leibrandt}}]{2019Hankin}%
  \BibitemOpen
  \bibfield  {author} {\bibinfo {author} {\bibfnamefont {A.~M.}\ \bibnamefont {Hankin}}, \bibinfo {author} {\bibfnamefont {E.~R.}\ \bibnamefont {Clements}}, \bibinfo {author} {\bibfnamefont {Y.}~\bibnamefont {Huang}}, \bibinfo {author} {\bibfnamefont {S.~M.}\ \bibnamefont {Brewer}}, \bibinfo {author} {\bibfnamefont {J.-S.}\ \bibnamefont {Chen}}, \bibinfo {author} {\bibfnamefont {C.-W.}\ \bibnamefont {Chou}}, \bibinfo {author} {\bibfnamefont {D.~B.}\ \bibnamefont {Hume}},\ and\ \bibinfo {author} {\bibfnamefont {D.~R.}\ \bibnamefont {Leibrandt}},\ }\bibfield  {title} {\bibinfo {title} {Systematic uncertainty due to background-gas collisions in trapped-ion optical clocks},\ }\href {https://doi.org/10.1103/PhysRevA.100.033419} {\bibfield  {journal} {\bibinfo  {journal} {Phys. Rev. A}\ }\textbf {\bibinfo {volume} {100}},\ \bibinfo {pages} {033419} (\bibinfo {year} {2019})}\BibitemShut {NoStop}%
\bibitem [{\citenamefont {Dunn}(1968)}]{dunn1968a}%
  \BibitemOpen
  \bibfield  {author} {\bibinfo {author} {\bibfnamefont {G.~H.}\ \bibnamefont {Dunn}},\ }\bibfield  {title} {\bibinfo {title} {Photodissociation of {H$_2^+$} and {D$_2^+$}: {Theory}},\ }\href {https://doi.org/10.1103/PhysRev.172.1} {\bibfield  {journal} {\bibinfo  {journal} {Phys. Rev.}\ }\textbf {\bibinfo {volume} {172}},\ \bibinfo {pages} {1} (\bibinfo {year} {1968})}\BibitemShut {NoStop}%
\bibitem [{\citenamefont {Karr}(2024)}]{PCKarr2024}%
  \BibitemOpen
  \bibfield  {author} {\bibinfo {author} {\bibfnamefont {J.-P.}\ \bibnamefont {Karr}}} (\bibinfo {year} {2024}),\ \bibinfo {note} {private communication: Photodissociation cross-section of $\mathrm{H}_2^+$ ($\nu=0,L=0$) at \SI{313}{nm}:\\ $\sigma= \SI{1.430E-26}{cm^2}$}\BibitemShut {NoStop}%
\bibitem [{\citenamefont {Karr}\ \emph {et~al.}(2008{\natexlab{a}})\citenamefont {Karr}, \citenamefont {Korobov},\ and\ \citenamefont {Hilico}}]{karr2008}%
  \BibitemOpen
  \bibfield  {author} {\bibinfo {author} {\bibfnamefont {J.-P.}\ \bibnamefont {Karr}}, \bibinfo {author} {\bibfnamefont {V.~I.}\ \bibnamefont {Korobov}},\ and\ \bibinfo {author} {\bibfnamefont {L.}~\bibnamefont {Hilico}},\ }\bibfield  {title} {\bibinfo {title} {Vibrational spectroscopy of {H$_2^+$}: {Precise} evaluation of the {Zeeman} effect},\ }\href {https://doi.org/10.1103/PhysRevA.77.062507} {\bibfield  {journal} {\bibinfo  {journal} {Phys. Rev. A}\ }\textbf {\bibinfo {volume} {77}},\ \bibinfo {pages} {062507} (\bibinfo {year} {2008}{\natexlab{a}})}\BibitemShut {NoStop}%
\bibitem [{\citenamefont {Stadler}(2023)}]{23Stadler}%
  \BibitemOpen
  \bibfield  {author} {\bibinfo {author} {\bibfnamefont {O.}~\bibnamefont {Stadler}},\ }\bibfield  {title} {\bibinfo {title} {State preparation and readout routine for molecular ions},\ }\href {https://ethz.ch/content/dam/ethz/special-interest/phys/quantum-electronics/tiqi-dam/documents/masters_theses/Masterthesis_Oliver_Stadler} {\bibfield  {journal} {\bibinfo  {journal} {Master's Thesis, ETH Z\"urich}\ } (\bibinfo {year} {2023})}\BibitemShut {NoStop}%
\bibitem [{\citenamefont {Ramsey}(1956)}]{ramsey1956}%
  \BibitemOpen
  \bibfield  {author} {\bibinfo {author} {\bibfnamefont {N.~F.}\ \bibnamefont {Ramsey}},\ }\href@noop {} {\emph {\bibinfo {title} {Molecular {Beams}}}}\ (\bibinfo  {publisher} {Oxford University Press},\ \bibinfo {year} {1956})\ Chap.\ \bibinfo {chapter} {V.3}\BibitemShut {NoStop}%
\bibitem [{\citenamefont {Hernández~Vera}\ \emph {et~al.}(2017)\citenamefont {Hernández~Vera}, \citenamefont {Gianturco}, \citenamefont {Wester}, \citenamefont {da~Silva}, \citenamefont {Dulieu},\ and\ \citenamefont {Schiller}}]{17Vera1}%
  \BibitemOpen
  \bibfield  {author} {\bibinfo {author} {\bibfnamefont {M.}~\bibnamefont {Hernández~Vera}}, \bibinfo {author} {\bibfnamefont {F.~A.}\ \bibnamefont {Gianturco}}, \bibinfo {author} {\bibfnamefont {R.}~\bibnamefont {Wester}}, \bibinfo {author} {\bibfnamefont {H.}~\bibnamefont {da~Silva}}, \bibinfo {author} {\bibfnamefont {O.}~\bibnamefont {Dulieu}},\ and\ \bibinfo {author} {\bibfnamefont {S.}~\bibnamefont {Schiller}},\ }\bibfield  {title} {\bibinfo {title} {Rotationally inelastic collisions of \ensuremath{\mathrm{\uppercase{h}}_2^+} ions with \ensuremath{\mathrm{\uppercase{h}e}} buffer gas: Computing cross sections and rates},\ }\href {https://doi.org/10.1063/1.4978475} {\bibfield  {journal} {\bibinfo  {journal} {The Journal of Chemical Physics}\ }\textbf {\bibinfo {volume} {146}},\ \bibinfo {pages} {124310} (\bibinfo {year} {2017})}\BibitemShut {NoStop}%
\bibitem [{\citenamefont {Richardson}\ \emph {et~al.}(1968)\citenamefont {Richardson}, \citenamefont {Jefferts},\ and\ \citenamefont {Dehmelt}}]{1968Richardson}%
  \BibitemOpen
  \bibfield  {author} {\bibinfo {author} {\bibfnamefont {C.}~\bibnamefont {Richardson}}, \bibinfo {author} {\bibfnamefont {K.-B.}\ \bibnamefont {Jefferts}},\ and\ \bibinfo {author} {\bibfnamefont {H.}~\bibnamefont {Dehmelt}},\ }\bibfield  {title} {\bibinfo {title} {Alignment of the {H}$_2^+$ molecular ion by selective photodissociation. ii. experiments on the radio-frequency spectrum},\ }\href {https://doi.org/10.1103/PhysRev.165.80} {\bibfield  {journal} {\bibinfo  {journal} {Physical Review}\ }\textbf {\bibinfo {volume} {165}},\ \bibinfo {pages} {80} (\bibinfo {year} {1968})}\BibitemShut {NoStop}%
\bibitem [{\citenamefont {Jefferts}(1969)}]{1969Jefferts}%
  \BibitemOpen
  \bibfield  {author} {\bibinfo {author} {\bibfnamefont {K.~B.}\ \bibnamefont {Jefferts}},\ }\bibfield  {title} {\bibinfo {title} {Hyperfine structure in the molecular ion {H}$_2^+$},\ }\href {https://doi.org/10.1103/PhysRevLett.23.1476} {\bibfield  {journal} {\bibinfo  {journal} {Physical Review Letters}\ }\textbf {\bibinfo {volume} {23}},\ \bibinfo {pages} {1476} (\bibinfo {year} {1969})}\BibitemShut {NoStop}%
\bibitem [{\citenamefont {Fu}\ \emph {et~al.}(1992)\citenamefont {Fu}, \citenamefont {Hessels},\ and\ \citenamefont {Lundeen}}]{1992Fu}%
  \BibitemOpen
  \bibfield  {author} {\bibinfo {author} {\bibfnamefont {Z.~W.}\ \bibnamefont {Fu}}, \bibinfo {author} {\bibfnamefont {E.~A.}\ \bibnamefont {Hessels}},\ and\ \bibinfo {author} {\bibfnamefont {S.~R.}\ \bibnamefont {Lundeen}},\ }\bibfield  {title} {\bibinfo {title} {Determination of the hyperfine structure of {H}$_2^+$(\ensuremath{\nu}=0, {R}=1) by microwave spectroscopy of high-{L} n=27 {R}ydberg states of {H}$_2$},\ }\href {https://doi.org/10.1103/PhysRevA.46.R5313} {\bibfield  {journal} {\bibinfo  {journal} {Phys. Rev. A}\ }\textbf {\bibinfo {volume} {46}},\ \bibinfo {pages} {R5313} (\bibinfo {year} {1992})}\BibitemShut {NoStop}%
\bibitem [{\citenamefont {Osterwalder}\ \emph {et~al.}(2004)\citenamefont {Osterwalder}, \citenamefont {W\"uest}, \citenamefont {Merkt},\ and\ \citenamefont {Jungen}}]{2004Osterwalder}%
  \BibitemOpen
  \bibfield  {author} {\bibinfo {author} {\bibfnamefont {A.}~\bibnamefont {Osterwalder}}, \bibinfo {author} {\bibfnamefont {A.}~\bibnamefont {W\"uest}}, \bibinfo {author} {\bibfnamefont {F.}~\bibnamefont {Merkt}},\ and\ \bibinfo {author} {\bibfnamefont {C.}~\bibnamefont {Jungen}},\ }\bibfield  {title} {\bibinfo {title} {High-resolution millimeter wave spectroscopy and multichannel quantum defect theory of the hyperfine structure in high {R}ydberg states of molecular hydrogen {H$_{2}$}},\ }\href {https://doi.org/10.1063/1.1792596} {\bibfield  {journal} {\bibinfo  {journal} {The Journal of Chemical Physics}\ }\textbf {\bibinfo {volume} {121}},\ \bibinfo {pages} {11810} (\bibinfo {year} {2004})}\BibitemShut {NoStop}%
\bibitem [{\citenamefont {Menasian}\ and\ \citenamefont {Dehmelt}(1973)}]{1973Menasian}%
  \BibitemOpen
  \bibfield  {author} {\bibinfo {author} {\bibfnamefont {S.~C.}\ \bibnamefont {Menasian}}\ and\ \bibinfo {author} {\bibfnamefont {H.~G.}\ \bibnamefont {Dehmelt}},\ }\bibfield  {title} {\bibinfo {title} {High resolution study of the (1, 1/2, 1/2) -- (1, 1/2, 3/2) {Hfs} transitions in $\mathrm{H}_2^+$},\ }\href@noop {} {\bibfield  {journal} {\bibinfo  {journal} {Bull. Am. Phys. Soc.}\ }\textbf {\bibinfo {volume} {18}},\ \bibinfo {pages} {408} (\bibinfo {year} {1973})}\BibitemShut {NoStop}%
\bibitem [{\citenamefont {Menasian}(1973)}]{ThMenasian}%
  \BibitemOpen
  \bibfield  {author} {\bibinfo {author} {\bibfnamefont {S.~C.}\ \bibnamefont {Menasian}},\ }\bibfield  {title} {\bibinfo {title} {High resolution study of the $(\mathrm{F} \, \mathrm{F}_2) = (3/2 \, 1/2) \rightarrow (1/2 \, 1/2)$ {HFS} transition in stored $\mathrm{H}_2^+$ molecular ions},\ }\href@noop {} {\bibfield  {journal} {\bibinfo  {journal} {PhD thesis, University of Washington}\ } (\bibinfo {year} {1973})}\BibitemShut {NoStop}%
\bibitem [{\citenamefont {Haidar}\ \emph {et~al.}(2022)\citenamefont {Haidar}, \citenamefont {Korobov}, \citenamefont {Hilico},\ and\ \citenamefont {Karr}}]{haidar2022}%
  \BibitemOpen
  \bibfield  {author} {\bibinfo {author} {\bibfnamefont {M.}~\bibnamefont {Haidar}}, \bibinfo {author} {\bibfnamefont {V.~I.}\ \bibnamefont {Korobov}}, \bibinfo {author} {\bibfnamefont {L.}~\bibnamefont {Hilico}},\ and\ \bibinfo {author} {\bibfnamefont {J.-P.}\ \bibnamefont {Karr}},\ }\bibfield  {title} {\bibinfo {title} {Higher-order corrections to spin-orbit and spin-spin tensor interactions in hydrogen molecular ions: Theory and application to \ensuremath{\mathrm{\uppercase{h}}_2^+}},\ }\href {https://doi.org/10.1103/PhysRevA.106.022816} {\bibfield  {journal} {\bibinfo  {journal} {Phys. Rev. A}\ }\textbf {\bibinfo {volume} {106}},\ \bibinfo {pages} {022816} (\bibinfo {year} {2022})}\BibitemShut {NoStop}%
\bibitem [{\citenamefont {Myers}(2018)}]{18Myers}%
  \BibitemOpen
  \bibfield  {author} {\bibinfo {author} {\bibfnamefont {E.~G.}\ \bibnamefont {Myers}},\ }\bibfield  {title} {\bibinfo {title} {{C}{P}{T} tests with the antihydrogen molecular ion},\ }\href {https://doi.org/10.1103/PhysRevA.98.010101} {\bibfield  {journal} {\bibinfo  {journal} {Phys. Rev. A}\ }\textbf {\bibinfo {volume} {98}},\ \bibinfo {pages} {010101} (\bibinfo {year} {2018})}\BibitemShut {NoStop}%
\bibitem [{\citenamefont {Schiller}\ and\ \citenamefont {Karr}(2024{\natexlab{b}})}]{2024Schiller}%
  \BibitemOpen
  \bibfield  {author} {\bibinfo {author} {\bibfnamefont {S.}~\bibnamefont {Schiller}}\ and\ \bibinfo {author} {\bibfnamefont {J.-P.}\ \bibnamefont {Karr}},\ }\bibfield  {title} {\bibinfo {title} {Prospects for the determination of fundamental constants with beyond-state-of-the-art uncertainty using molecular hydrogen ion spectroscopy},\ }\href {https://doi.org/10.1103/PhysRevA.109.042825} {\bibfield  {journal} {\bibinfo  {journal} {Phys. Rev. A}\ }\textbf {\bibinfo {volume} {109}},\ \bibinfo {pages} {042825} (\bibinfo {year} {2024}{\natexlab{b}})}\BibitemShut {NoStop}%
\bibitem [{\citenamefont {Danev}\ \emph {et~al.}(2021)\citenamefont {Danev}, \citenamefont {Bakalov}, \citenamefont {Korobov},\ and\ \citenamefont {Schiller}}]{2021Danev}%
  \BibitemOpen
  \bibfield  {author} {\bibinfo {author} {\bibfnamefont {P.}~\bibnamefont {Danev}}, \bibinfo {author} {\bibfnamefont {D.}~\bibnamefont {Bakalov}}, \bibinfo {author} {\bibfnamefont {V.~I.}\ \bibnamefont {Korobov}},\ and\ \bibinfo {author} {\bibfnamefont {S.}~\bibnamefont {Schiller}},\ }\bibfield  {title} {\bibinfo {title} {Hyperfine structure and electric quadrupole transitions in the deuterium molecular ion},\ }\href {https://doi.org/10.1103/PhysRevA.103.012805} {\bibfield  {journal} {\bibinfo  {journal} {Phys. Rev. A}\ }\textbf {\bibinfo {volume} {103}},\ \bibinfo {pages} {012805} (\bibinfo {year} {2021})}\BibitemShut {NoStop}%
\bibitem [{\citenamefont {K\"onig}\ \emph {et~al.}(2025)\citenamefont {K\"onig}, \citenamefont {Hei\ss{}e}, \citenamefont {Morgner}, \citenamefont {Sailer}, \citenamefont {Tu}, \citenamefont {Bakalov}, \citenamefont {Blaum}, \citenamefont {Schiller},\ and\ \citenamefont {Sturm}}]{25konig}%
  \BibitemOpen
  \bibfield  {author} {\bibinfo {author} {\bibfnamefont {C.~M.}\ \bibnamefont {K\"onig}}, \bibinfo {author} {\bibfnamefont {F.}~\bibnamefont {Hei\ss{}e}}, \bibinfo {author} {\bibfnamefont {J.}~\bibnamefont {Morgner}}, \bibinfo {author} {\bibfnamefont {T.}~\bibnamefont {Sailer}}, \bibinfo {author} {\bibfnamefont {B.}~\bibnamefont {Tu}}, \bibinfo {author} {\bibfnamefont {D.}~\bibnamefont {Bakalov}}, \bibinfo {author} {\bibfnamefont {K.}~\bibnamefont {Blaum}}, \bibinfo {author} {\bibfnamefont {S.}~\bibnamefont {Schiller}},\ and\ \bibinfo {author} {\bibfnamefont {S.}~\bibnamefont {Sturm}},\ }\bibfield  {title} {\bibinfo {title} {Nondestructive control of the rovibrational ground state of a single molecular hydrogen ion in a penning trap},\ }\href {https://doi.org/10.1103/PhysRevLett.134.163001} {\bibfield  {journal} {\bibinfo  {journal} {Phys. Rev. Lett.}\ }\textbf {\bibinfo {volume} {134}},\ \bibinfo {pages} {163001} (\bibinfo {year} {2025})}\BibitemShut {NoStop}%
\bibitem [{\citenamefont {W\"ubbena}\ \emph {et~al.}(2012)\citenamefont {W\"ubbena}, \citenamefont {Amairi}, \citenamefont {Mandel},\ and\ \citenamefont {Schmidt}}]{12Wubbena}%
  \BibitemOpen
  \bibfield  {author} {\bibinfo {author} {\bibfnamefont {J.~B.}\ \bibnamefont {W\"ubbena}}, \bibinfo {author} {\bibfnamefont {S.}~\bibnamefont {Amairi}}, \bibinfo {author} {\bibfnamefont {O.}~\bibnamefont {Mandel}},\ and\ \bibinfo {author} {\bibfnamefont {P.~O.}\ \bibnamefont {Schmidt}},\ }\bibfield  {title} {\bibinfo {title} {Sympathetic cooling of mixed-species two-ion crystals for precision spectroscopy},\ }\href {https://doi.org/10.1103/PhysRevA.85.043412} {\bibfield  {journal} {\bibinfo  {journal} {Phys. Rev. A}\ }\textbf {\bibinfo {volume} {85}},\ \bibinfo {pages} {043412} (\bibinfo {year} {2012})}\BibitemShut {NoStop}%
\bibitem [{\citenamefont {Schwegler}(2024)}]{theswagler}%
  \BibitemOpen
  \bibfield  {author} {\bibinfo {author} {\bibfnamefont {N.}~\bibnamefont {Schwegler}},\ }\bibfield  {title} {\bibinfo {title} {Quantum logic detection of a single $\mathrm{H}_2^+$ ion},\ }\href {https://doi.org/10.3929/ethz-b-000699389} {\bibfield  {journal} {\bibinfo  {journal} {PhD thesis, ETH Z\"urich}\ } (\bibinfo {year} {2024})}\BibitemShut {NoStop}%
\bibitem [{\citenamefont {Barakhshan}\ \emph {et~al.}()\citenamefont {Barakhshan}, \citenamefont {Marrs}, \citenamefont {Bhosale}, \citenamefont {Arora}, \citenamefont {Eigenmann},\ and\ \citenamefont {Safronova}}]{22UDportal}%
  \BibitemOpen
  \bibfield  {author} {\bibinfo {author} {\bibfnamefont {P.}~\bibnamefont {Barakhshan}}, \bibinfo {author} {\bibfnamefont {A.}~\bibnamefont {Marrs}}, \bibinfo {author} {\bibfnamefont {A.}~\bibnamefont {Bhosale}}, \bibinfo {author} {\bibfnamefont {B.}~\bibnamefont {Arora}}, \bibinfo {author} {\bibfnamefont {R.}~\bibnamefont {Eigenmann}},\ and\ \bibinfo {author} {\bibfnamefont {M.~S.}\ \bibnamefont {Safronova}},\ }\href {https://www.udel.edu/atom} {\bibinfo {title} {\textit{Portal for High-Precision Atomic Data and Computation} (version 2.0) (2022)}}\BibitemShut {NoStop}%
\bibitem [{\citenamefont {Wineland}\ and\ \citenamefont {Dehmelt}(1975)}]{1975WinelandAxial}%
  \BibitemOpen
  \bibfield  {author} {\bibinfo {author} {\bibfnamefont {D.}~\bibnamefont {Wineland}}\ and\ \bibinfo {author} {\bibfnamefont {H.}~\bibnamefont {Dehmelt}},\ }\bibfield  {title} {\bibinfo {title} {Line shifts and widths of axial, cyclotron and g-2 resonances in tailored, stored electron (ion) cloud},\ }\href {https://doi.org/https://doi.org/10.1016/0020-7381(75)87031-8} {\bibfield  {journal} {\bibinfo  {journal} {International Journal of Mass Spectrometry and Ion Physics}\ }\textbf {\bibinfo {volume} {16}},\ \bibinfo {pages} {338} (\bibinfo {year} {1975})}\BibitemShut {NoStop}%
\bibitem [{\citenamefont {Gorman}\ \emph {et~al.}(2014)\citenamefont {Gorman}, \citenamefont {Schindler}, \citenamefont {Selvarajan}, \citenamefont {Daniilidis},\ and\ \citenamefont {H\"affner}}]{2014Gorman}%
  \BibitemOpen
  \bibfield  {author} {\bibinfo {author} {\bibfnamefont {D.~J.}\ \bibnamefont {Gorman}}, \bibinfo {author} {\bibfnamefont {P.}~\bibnamefont {Schindler}}, \bibinfo {author} {\bibfnamefont {S.}~\bibnamefont {Selvarajan}}, \bibinfo {author} {\bibfnamefont {N.}~\bibnamefont {Daniilidis}},\ and\ \bibinfo {author} {\bibfnamefont {H.}~\bibnamefont {H\"affner}},\ }\bibfield  {title} {\bibinfo {title} {Two-mode coupling in a single-ion oscillator via parametric resonance},\ }\href {https://doi.org/10.1103/PhysRevA.89.062332} {\bibfield  {journal} {\bibinfo  {journal} {Phys. Rev. A}\ }\textbf {\bibinfo {volume} {89}},\ \bibinfo {pages} {062332} (\bibinfo {year} {2014})}\BibitemShut {NoStop}%
\bibitem [{\citenamefont {Hou}\ \emph {et~al.}(2024)\citenamefont {Hou}, \citenamefont {Wu}, \citenamefont {Erickson}, \citenamefont {Zarantonello}, \citenamefont {Brandt}, \citenamefont {Cole}, \citenamefont {Wilson}, \citenamefont {Slichter},\ and\ \citenamefont {Leibfried}}]{2024Hou}%
  \BibitemOpen
  \bibfield  {author} {\bibinfo {author} {\bibfnamefont {P.-Y.}\ \bibnamefont {Hou}}, \bibinfo {author} {\bibfnamefont {J.~J.}\ \bibnamefont {Wu}}, \bibinfo {author} {\bibfnamefont {S.~D.}\ \bibnamefont {Erickson}}, \bibinfo {author} {\bibfnamefont {G.}~\bibnamefont {Zarantonello}}, \bibinfo {author} {\bibfnamefont {A.~D.}\ \bibnamefont {Brandt}}, \bibinfo {author} {\bibfnamefont {D.~C.}\ \bibnamefont {Cole}}, \bibinfo {author} {\bibfnamefont {A.~C.}\ \bibnamefont {Wilson}}, \bibinfo {author} {\bibfnamefont {D.~H.}\ \bibnamefont {Slichter}},\ and\ \bibinfo {author} {\bibfnamefont {D.}~\bibnamefont {Leibfried}},\ }\bibfield  {title} {\bibinfo {title} {Indirect cooling of weakly coupled trapped-ion mechanical oscillators},\ }\href {https://doi.org/10.1103/PhysRevX.14.021003} {\bibfield  {journal} {\bibinfo  {journal} {Phys. Rev. X}\ }\textbf {\bibinfo {volume} {14}},\ \bibinfo {pages} {021003} (\bibinfo {year} {2024})}\BibitemShut {NoStop}%
\bibitem [{\citenamefont {King}\ \emph {et~al.}(2021)\citenamefont {King}, \citenamefont {Spie\ss{}}, \citenamefont {Micke}, \citenamefont {Wilzewski}, \citenamefont {Leopold}, \citenamefont {Crespo L\'opez-Urrutia},\ and\ \citenamefont {Schmidt}}]{2021King}%
  \BibitemOpen
  \bibfield  {author} {\bibinfo {author} {\bibfnamefont {S.~A.}\ \bibnamefont {King}}, \bibinfo {author} {\bibfnamefont {L.~J.}\ \bibnamefont {Spie\ss{}}}, \bibinfo {author} {\bibfnamefont {P.}~\bibnamefont {Micke}}, \bibinfo {author} {\bibfnamefont {A.}~\bibnamefont {Wilzewski}}, \bibinfo {author} {\bibfnamefont {T.}~\bibnamefont {Leopold}}, \bibinfo {author} {\bibfnamefont {J.~R.}\ \bibnamefont {Crespo L\'opez-Urrutia}},\ and\ \bibinfo {author} {\bibfnamefont {P.~O.}\ \bibnamefont {Schmidt}},\ }\bibfield  {title} {\bibinfo {title} {Algorithmic ground-state cooling of weakly coupled oscillators using quantum logic},\ }\href {https://doi.org/10.1103/PhysRevX.11.041049} {\bibfield  {journal} {\bibinfo  {journal} {Phys. Rev. X}\ }\textbf {\bibinfo {volume} {11}},\ \bibinfo {pages} {041049} (\bibinfo {year} {2021})}\BibitemShut {NoStop}%
\bibitem [{\citenamefont {Korobov}\ \emph {et~al.}(2006)\citenamefont {Korobov}, \citenamefont {Hilico},\ and\ \citenamefont {Karr}}]{korobov2006}%
  \BibitemOpen
  \bibfield  {author} {\bibinfo {author} {\bibfnamefont {V.~I.}\ \bibnamefont {Korobov}}, \bibinfo {author} {\bibfnamefont {L.}~\bibnamefont {Hilico}},\ and\ \bibinfo {author} {\bibfnamefont {J.-P.}\ \bibnamefont {Karr}},\ }\bibfield  {title} {\bibinfo {title} {Hyperfine structure in the hydrogen molecular ion},\ }\href {https://doi.org/10.1103/PhysRevA.74.040502} {\bibfield  {journal} {\bibinfo  {journal} {Phys. Rev. A}\ }\textbf {\bibinfo {volume} {74}},\ \bibinfo {pages} {040502} (\bibinfo {year} {2006})}\BibitemShut {NoStop}%
\bibitem [{\citenamefont {Karr}\ \emph {et~al.}(2008{\natexlab{b}})\citenamefont {Karr}, \citenamefont {Bielsa}, \citenamefont {Douillet}, \citenamefont {Pedregosa~Gutierrez}, \citenamefont {Korobov},\ and\ \citenamefont {Hilico}}]{08Karr}%
  \BibitemOpen
  \bibfield  {author} {\bibinfo {author} {\bibfnamefont {J.-P.}\ \bibnamefont {Karr}}, \bibinfo {author} {\bibfnamefont {F.}~\bibnamefont {Bielsa}}, \bibinfo {author} {\bibfnamefont {A.}~\bibnamefont {Douillet}}, \bibinfo {author} {\bibfnamefont {J.}~\bibnamefont {Pedregosa~Gutierrez}}, \bibinfo {author} {\bibfnamefont {V.~I.}\ \bibnamefont {Korobov}},\ and\ \bibinfo {author} {\bibfnamefont {L.}~\bibnamefont {Hilico}},\ }\bibfield  {title} {\bibinfo {title} {Vibrational spectroscopy of \ensuremath{\mathrm{\uppercase{h}}_2^+}: Hyperfine structure of two-photon transitions},\ }\href {https://doi.org/10.1103/PhysRevA.77.063410} {\bibfield  {journal} {\bibinfo  {journal} {Phys. Rev. A}\ }\textbf {\bibinfo {volume} {77}},\ \bibinfo {pages} {063410} (\bibinfo {year} {2008}{\natexlab{b}})}\BibitemShut {NoStop}%
\bibitem [{\citenamefont {Karr}\ \emph {et~al.}(2020)\citenamefont {Karr}, \citenamefont {Haidar}, \citenamefont {Hilico}, \citenamefont {Zhong},\ and\ \citenamefont {Korobov}}]{2020KarrPRA}%
  \BibitemOpen
  \bibfield  {author} {\bibinfo {author} {\bibfnamefont {J.-P.}\ \bibnamefont {Karr}}, \bibinfo {author} {\bibfnamefont {M.}~\bibnamefont {Haidar}}, \bibinfo {author} {\bibfnamefont {L.}~\bibnamefont {Hilico}}, \bibinfo {author} {\bibfnamefont {Z.-X.}\ \bibnamefont {Zhong}},\ and\ \bibinfo {author} {\bibfnamefont {V.~I.}\ \bibnamefont {Korobov}},\ }\bibfield  {title} {\bibinfo {title} {Higher-order corrections to spin-spin scalar interactions in \ensuremath{\mathrm{\uppercase{hd}}^+} and \ensuremath{\mathrm{\uppercase{h}}_2^+}},\ }\href {https://doi.org/10.1103/PhysRevA.102.052827} {\bibfield  {journal} {\bibinfo  {journal} {Phys. Rev. A}\ }\textbf {\bibinfo {volume} {102}},\ \bibinfo {pages} {052827} (\bibinfo {year} {2020})}\BibitemShut {NoStop}%
\end{thebibliography}%
\newpage
\clearpage
\pagebreak

\renewcommand{\bibliography}[1]{} 

\end{document}